\documentclass[3p,times]{elsarticle}

\usepackage{amssymb}
\usepackage{lineno}
\usepackage{natbib}
\usepackage{graphicx}
\usepackage{array,multirow,makecell}
\usepackage{color}
\usepackage[figuresright]{rotating}

\begin{document}

\begin{frontmatter}




\title{A warm or a cold early Earth? New insights from a 3-D climate-carbon model}

\author[label1,label2]{Benjamin Charnay}
\author[label3]{Guillaume Le Hir}
\author[label3]{Fr\'ed\'eric Fluteau}
\author[label4]{Fran\c{c}ois Forget}
\author[label2,label5]{David C. Catling}

\address[label1]{LESIA, Observatoire de Paris, PSL Research University, CNRS, Sorbonne Universit\'es, UPMC Univ. Paris 06, Univ. Paris Diderot, Sorbonne Paris Cit\'e, 5 Place Jules Janssen, 92195 Meudon, France }
\address[label2]{Virtual Planetary Laboratory, University of Washington,
    Seattle, WA 98125, USA.}

\address[label3]{Institut de Physique du Globe de Paris, Sorbonne Paris Cité, Université Paris Diderot, 75238 Paris cedex 05, France}
\address[label4]{Laboratoire de M\'et\'eorologie Dynamique, IPSL/CNRS/UPMC, Paris 75005, France.}
\address[label5]{Department of Earth and Space Science, University of Washington, Seattle, WA 98125, USA.}

\begin{abstract}
Oxygen isotopes in marine cherts have been used to infer hot oceans during the Archean with temperatures between 60$^\circ$C (333 K) and 80$^\circ$C (353 K). Such climates are challenging for the early Earth warmed by the faint young Sun. The interpretation of the data has therefore been controversial. 1D climate modeling inferred that such hot climates would require very high levels of CO$_2$ (2-6 bars). Previous carbon cycle modeling concluded that such stable hot climates were impossible and that the carbon cycle should lead to cold climates during the Hadean and the Archean.
Here, we revisit the climate and carbon cycle of the early Earth at 3.8 Ga using a 3D climate-carbon model. We find that CO$_2$ partial pressures of around 1 bar could have produced hot climates given a low land fraction and cloud feedback effects. However, such high CO$_2$ partial pressures should not have been stable because of the weathering of terrestrial and oceanic basalts, producing an efficient stabilizing feedback. 
Moreover, the weathering of impact ejecta during the Late Heavy Bombardment (LHB) would have strongly reduced the CO$_2$ partial pressure leading to cold climates and potentially snowball Earth events after large impacts.
Our results therefore favor cold or temperate climates with global mean temperatures between around 8$^\circ$C (281 K) and 30$^\circ$C (303 K) and with 0.1-0.36 bar of CO$_2$ for the late Hadean and early Archean. 
Finally, our model suggests that the carbon cycle was efficient for preserving clement conditions on the early Earth without necessarily requiring any other greenhouse gas or warming process.
\end{abstract}

\begin{keyword}
early Earth, climate, carbon cycle, Hadean, Archean, Late Heavy Bombardment

\end{keyword}

\end{frontmatter}



\paragraph{}
\section{Introduction}

Life likely emerged on Earth before 3.5 Ga, during the Hadean or the early Archean \citep{buick81, nisbet01, bell15}.
Studying the atmosphere and the climates of the Earth during its first billion year is critical for understanding the environment in which life emerged and developed.
Oxygen isotopic ratio of Archean marine cherts have been interpreted to suggest hot oceans with temperatures between 60$^\circ$C (333 K) and 80$^\circ$C (353 K) \citep{knauth03, robert06}, compatible with the thermophiles inferred from evolutionary models of the ancient life \citep{gaucher08} and possibly with the indications for a low ocean water viscosity \citep{fralick11}.
However, such an interpretation has been strongly debated \citep{kasting06a, vandenboorn07, marin-carbonne14, tartese16} and other analyses suggest temperate oceans with temperatures lower than 40$^\circ$C \citep{hren09, blake10}. Moreover, at 3.5 Ga, the presence of glacial rocks at 20-40$^\circ$ latitude implies temperatures below 20$^\circ$C \citep{dewit16}, at least episodically.

In addition, previous modeling of the carbon cycle on the early Earth by \cite{sleep01} and \cite{zahnle02} suggested that the Archean was cold unless another strong greenhouse gas was present. They also suggested that the Hadean was likely very cold because of the weathering of impact ejecta, particularly during the Late Heavy Bombardment (LHB).
Therefore, the temperature of the early oceans remains an open question.

Here, we use a 3D climate-carbon model with pCO$_2$ ranging from 0.01 to 1 bar in order to determine physically plausible climates of the late Hadean and early Archean Earth. In section 2, we describe the climate simulations. In section 3 and 4, we analyze the carbon cycle and the effect of the LHB with respect to carbon cycle responses. We finish with a conclusion in section 5.

\section{Climate modeling}
\subsection{Description of the model}

We simulated the atmosphere of the early Earth using the Generic LMDZ GCM (Global Climate Model). This model solves the primitive hydrostatic equations of meteorology using a finite difference dynamical core on an Arakawa C grid. It uses robust and general parameterizations in order to simulate planets very different from the present-day Earth. The Generic LMDZ GCM has already been used for studying the temperate and cold climates of the Archean Earth \citep{charnay13}, the climates of early Mars with high CO$_2$ pressures \citep{forget13}, and the runaway greenhouse effect on terrestrial planets with high amounts of water vapor \citep{leconte13b}. It is thus adapted for simulating the early Earth with high amounts of CO$_2$ and potentially hot and moist climates. 

The radiative scheme is based on the correlated-k method.
At a given pressure and temperature, correlated-k coefficients in the GCM are interpolated from a matrix of coefficients stored in a 7 $\times$ 9 temperature and log-pressure grid: $T$=100, 150, 200, 250, 300, 350, 400 K, $p$ = 10$^{-1}$, 10$^0$, 10$^1$, ..., 10$^7$ Pa. We used 36 spectral bands in the thermal infrared and 38 at solar wavelengths. Sixteen points were used for the g-space integration, where g is the cumulative distribution function of the absorption data for each band.

Simulations were performed with a horizontal resolution of 64$\times$48 (corresponding to resolutions of 3.75$^\circ$ latitude by 5.625$^\circ$ longitude) and with 25 vertical layers with the lowest midlayer level at 5 m and the top level at 0.5 hPa. 
For the cloud microphysics, we either fixed the radii of water cloud particles (e.g. 12 $\mu$m for liquid droplets and 35 $\mu$m for icy particles for present-day Earth) or we fixed the density of cloud condensation nuclei (CCN) by mass of air (e.g. 5$\times$10$^6$ particles/kg for liquid droplets and 2$\times$10$^4$particles/kg for icy clouds for present-day Earth). As in \cite{charnay13}, the oceanic transport and the sea ice formation were computed with the 2-layer oceanic model from \cite{codron12}.

To investigate the early Earth at 3.8 Ga, we used a solar constant of 1024 W/m$^2$ corresponding to 75$\%$ of the present value (1361 W/m$^2$).
We ran simulations with no land. This hypothesis is valid for small continental surface fractions (e.g. $<$30$\%$ of the present-day fraction or $<$10$\%$ of the total Earth's surface) as we assumed at this time \citep{flament08} (see also section 3.1), although the land fraction is a matter of dispute \citep{viehmann14}. Such small continental fraction should indeed have a negligible direct impact on the global climate.
We also assumed that Earth's rotation period was 14h. 
We used an atmospheric composition with 1 bar of nitrogen, a partial pressure of CO$_2$ (pCO$_2$) from 0.01 to 1 bar, and either no CH$_4$ or 2 mbar of CH$_4$, spanning the range of plausible methane concentrations estimated by a model for early ecosystems \citep{kharecha05}.  \cite{wordsworth17} showed that CO$_2$-CH$_4$ collision-induced absorption could have produced a strong warming on early Mars if the CH$_4$ mixing ratio was higher than 1$\%$. With our CH$_4$ mixing ratio of 0.1-0.2 $\%$, CO$_2$-CH$_4$ CIA produce a warming lower than 0.1 K with the 1D version of the model. We thus neglected CO$_2$-CH$_4$ CIA in our 3D simulations.

\begin{table}[h!]
\begin{tabular}{|m{1.3cm}|m{1.4cm}|m{1.6cm}|m{1.5cm}|m{2.0cm}|m{1.7cm}|m{1.7cm}|}
\hline
pCO$_2$ (bar)	& pCH$_4$ (mbar) 		&	Tsurf ($^\circ$C)	&	Planetary albedo  	&	Precipitation (mm/day)	&	SW forcing (W/m$^2$)	&LW forcing (W/m$^2$)	\\ \hline	
0.01  		& 0					&	-11.8 			&	0.4 				&	1.8    				&	-38.3 				&+24.5 			\\	
0.01  		& 2					&	2.7	 			&	0.33 			&	2.2     				&	-42.0				&+20.2 			\\
0.1			& 0					&	7.4  				&	0.34				&	2.8   				&	-47.1 				&+24.9 			\\	
0.1			& 2					&	21.3  			&	0.28				&	3.5 					&	-39.8 				&+18.1 			\\			
0.3			& 0					&	22.1  			&	0.31				&	3.6   				&	-42.4 				&+20.8 			\\
0.3			& 2					&	34.7  			&	0.26				&	4.4					&	-32.6 				&+13.8 			\\		
0.5			& 0					&	34.7  			&	0.27 			&	4.6     				&	-30.4 				&+16.0 			\\
0.5			& 2					&	49.9  			&	0.23 			&	5.1     				&	-22.8 				&+12.3 			\\
1.0			& 0					&	59.5  			&	0.24				&	5.7   				&	-19.4 				&+14.2 			\\
1.0			& 2					&	67.5  			&	0.23				&	6.1   				&	-20.3 				&+14.0 			\\	
\hline
\end{tabular}
\label{table1}
\caption{Global mean values from the GCM. Columns from left to rights are: pCO$_2$ (in bar), pCH$_4$ (in mbar), the mean surface temperature (in $^\circ$C), the mean planetary albedo, the mean precipitation rate (in mm/day), the shortwave cloud forcing (in W/m$^2$) and the longwave cloud forcing (in W/m$^2$).}
\end{table}

\begin{figure}[h!]
\begin{center} 
	\noindent\includegraphics[width=10cm]{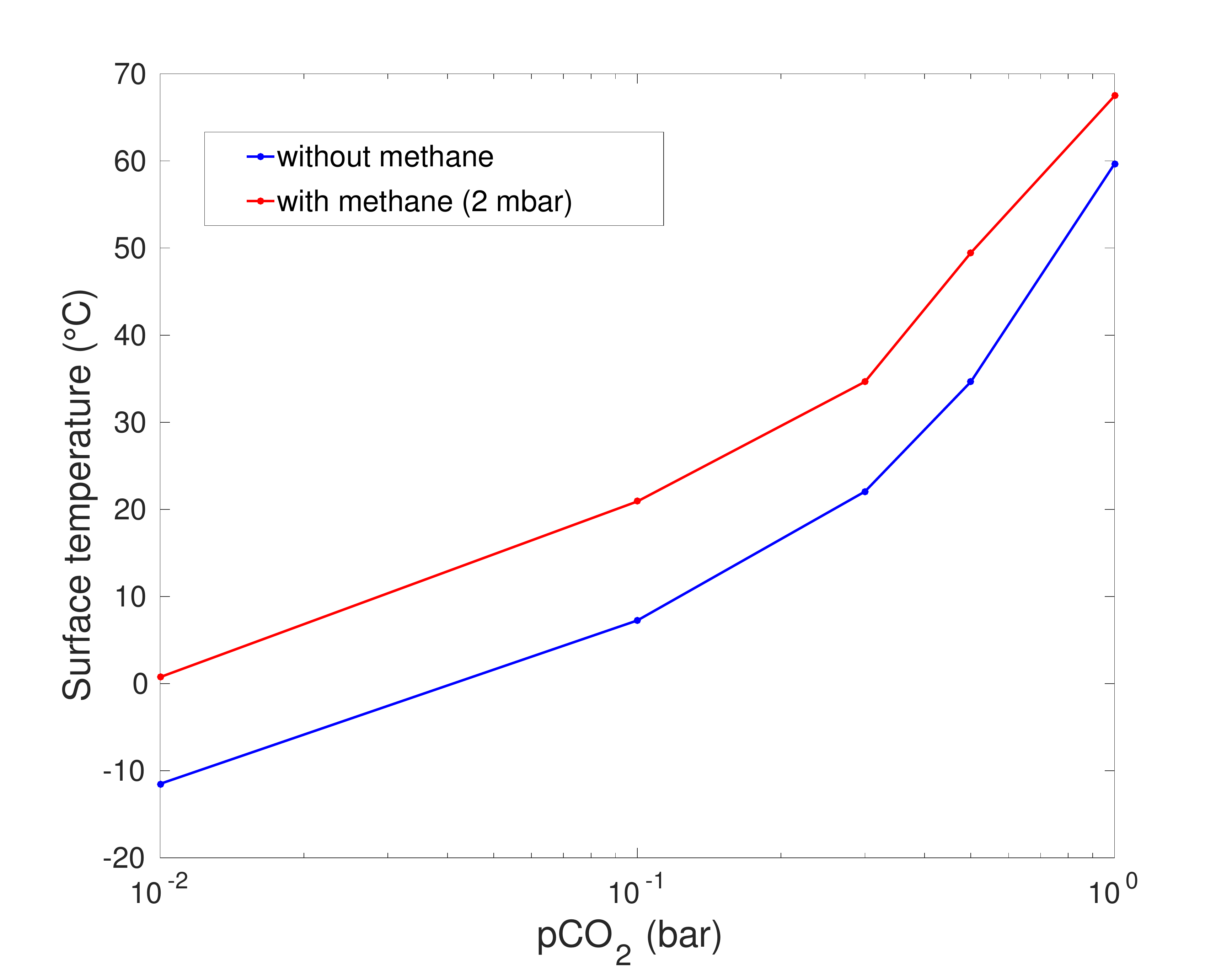}
	\noindent\includegraphics[width=10cm]{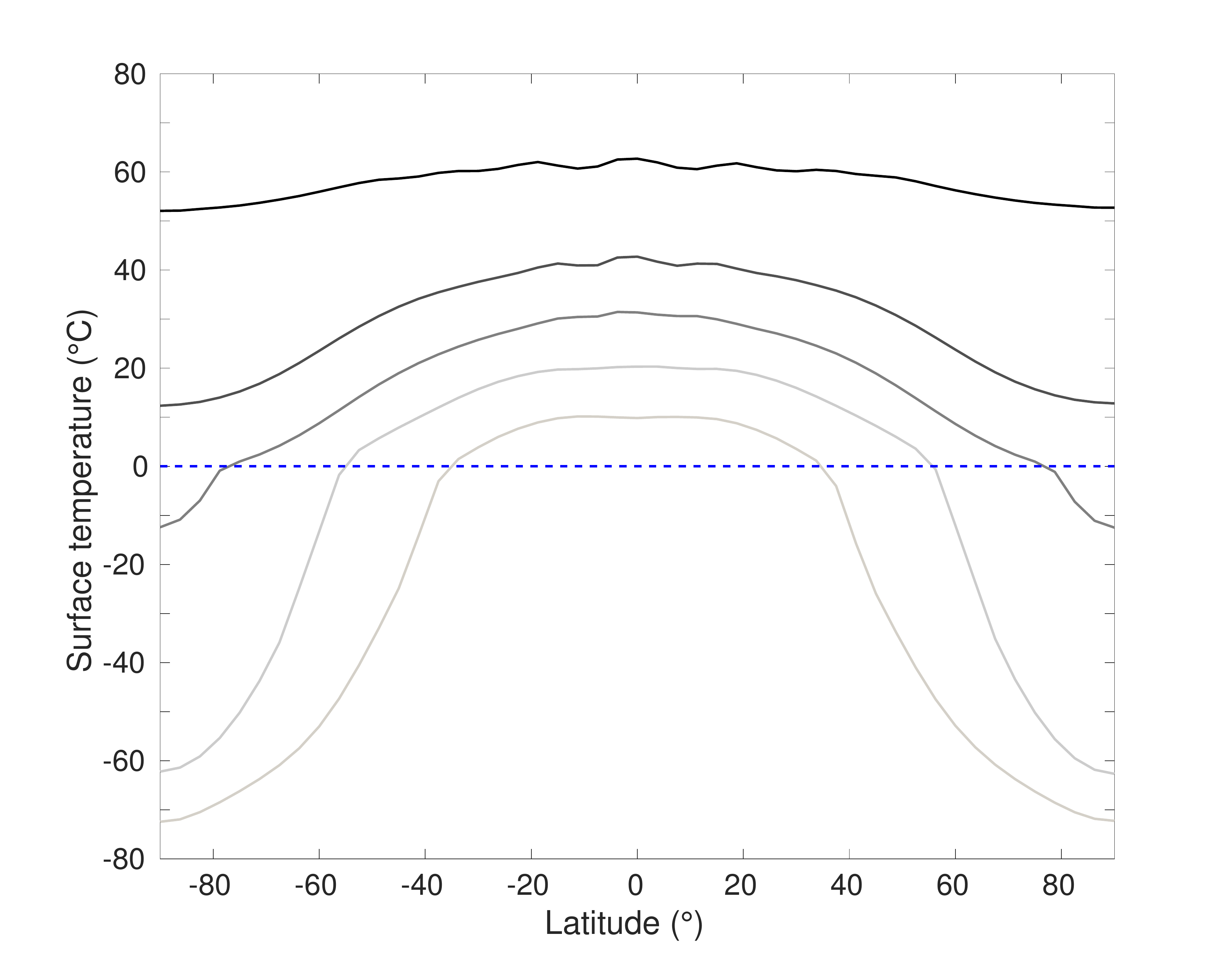}
\end{center} 
\caption{Top panel: global mean surface temperature without methane (blue line) and with 2 mbars of methane (red line).
Bottom panel: zonal mean surface temperature for an atmosphere with no methane and with 0.01, 0.1, 0.3, 0.5 and 1 bar of CO$_2$ for lines from light gray to black respectively. The blue dashed line is the freezing temperature of water.} 
\label{figure_1}
\end{figure} 

\begin{figure}[h!]
\begin{center} 
	\noindent\includegraphics[width=8cm]{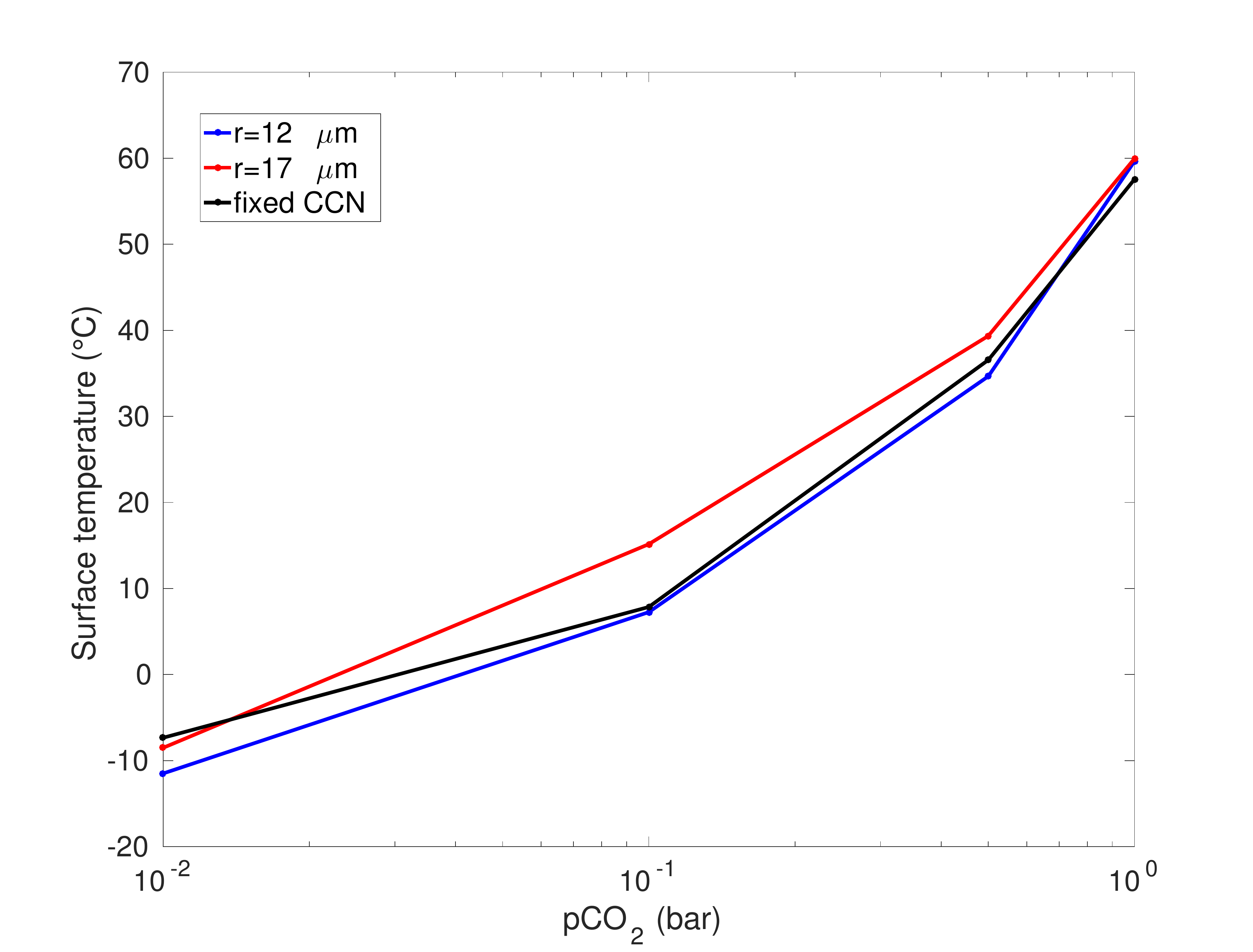}
	\noindent\includegraphics[width=8cm]{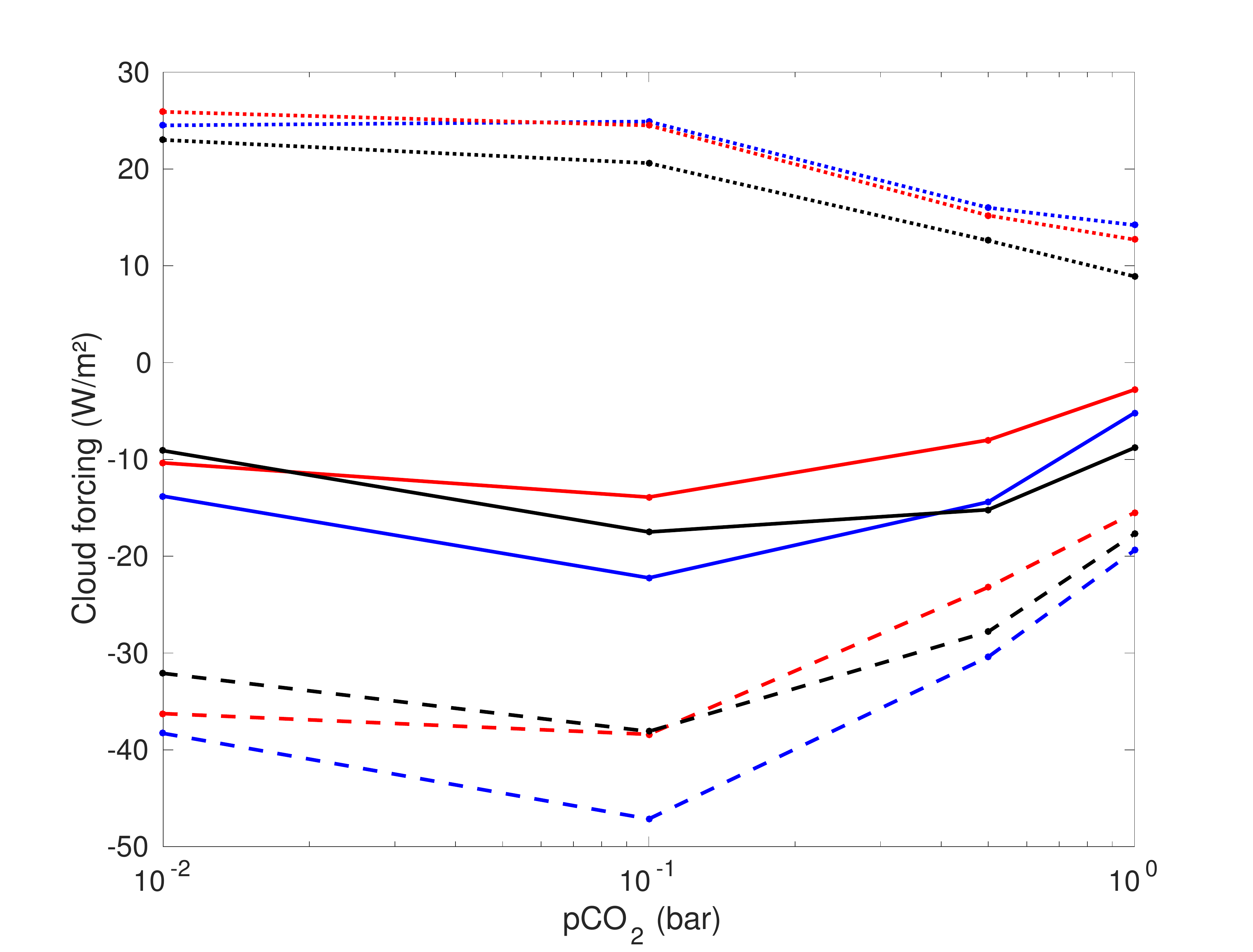}
\end{center} 
\caption{Left panel: global mean surface temperature with no methane with 12 $\mu$m liquid cloud droplets (blue), 17 $\mu$m liquid cloud droplets (red) and varying cloud droplet radii with a fixed CCN concentration (black line).
Right: shortwave cloud forcing (dotted lines), longwave cloud forcing (dashed lines) and net cloud forcing (solid lines). The colors correspond to the same cases as the left panel.
} 
\label{figure_2}
\end{figure}

\begin{figure}[h!] 
\begin{center} 
	\includegraphics[width=5.cm]{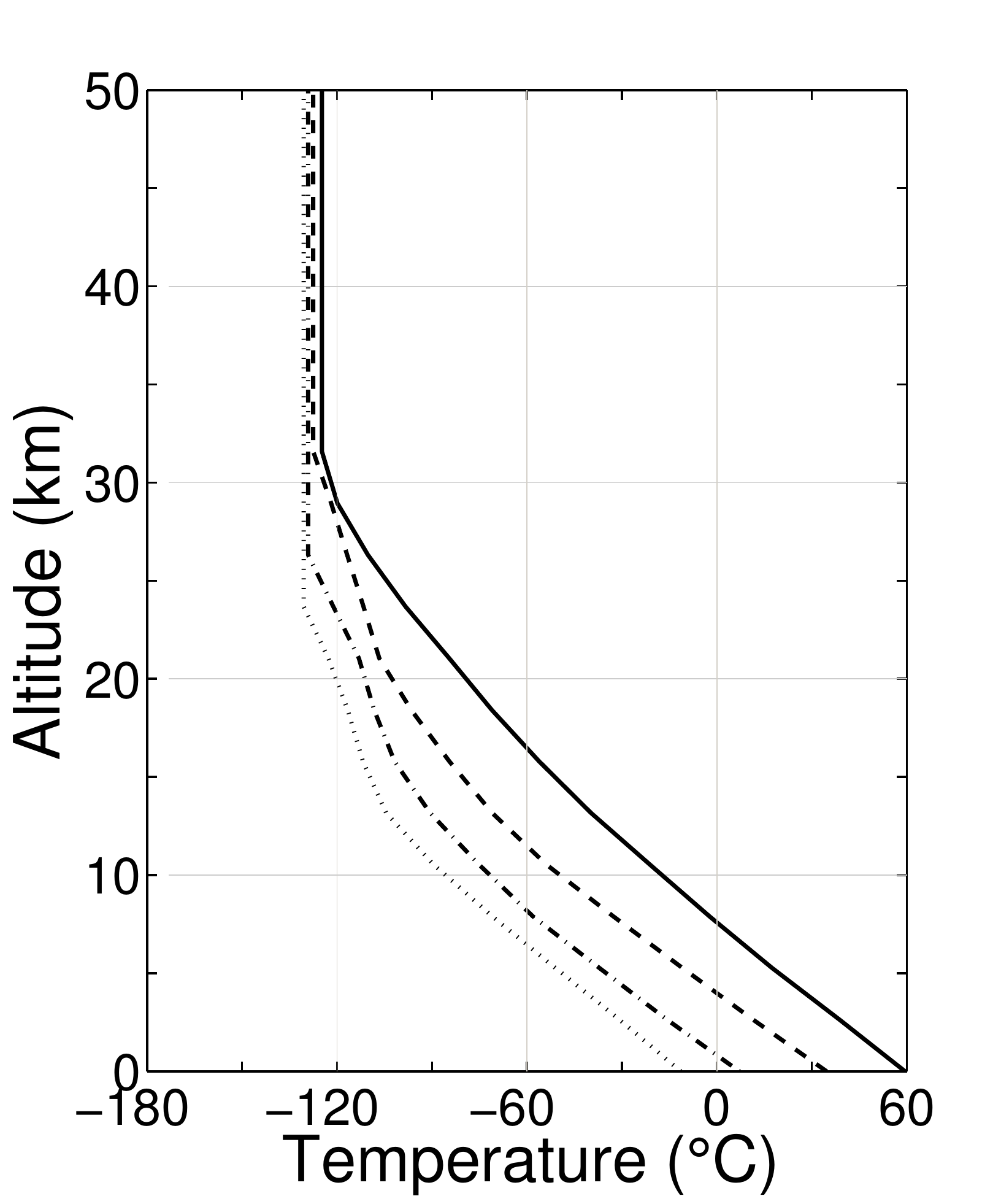}
	\includegraphics[width=5.cm]{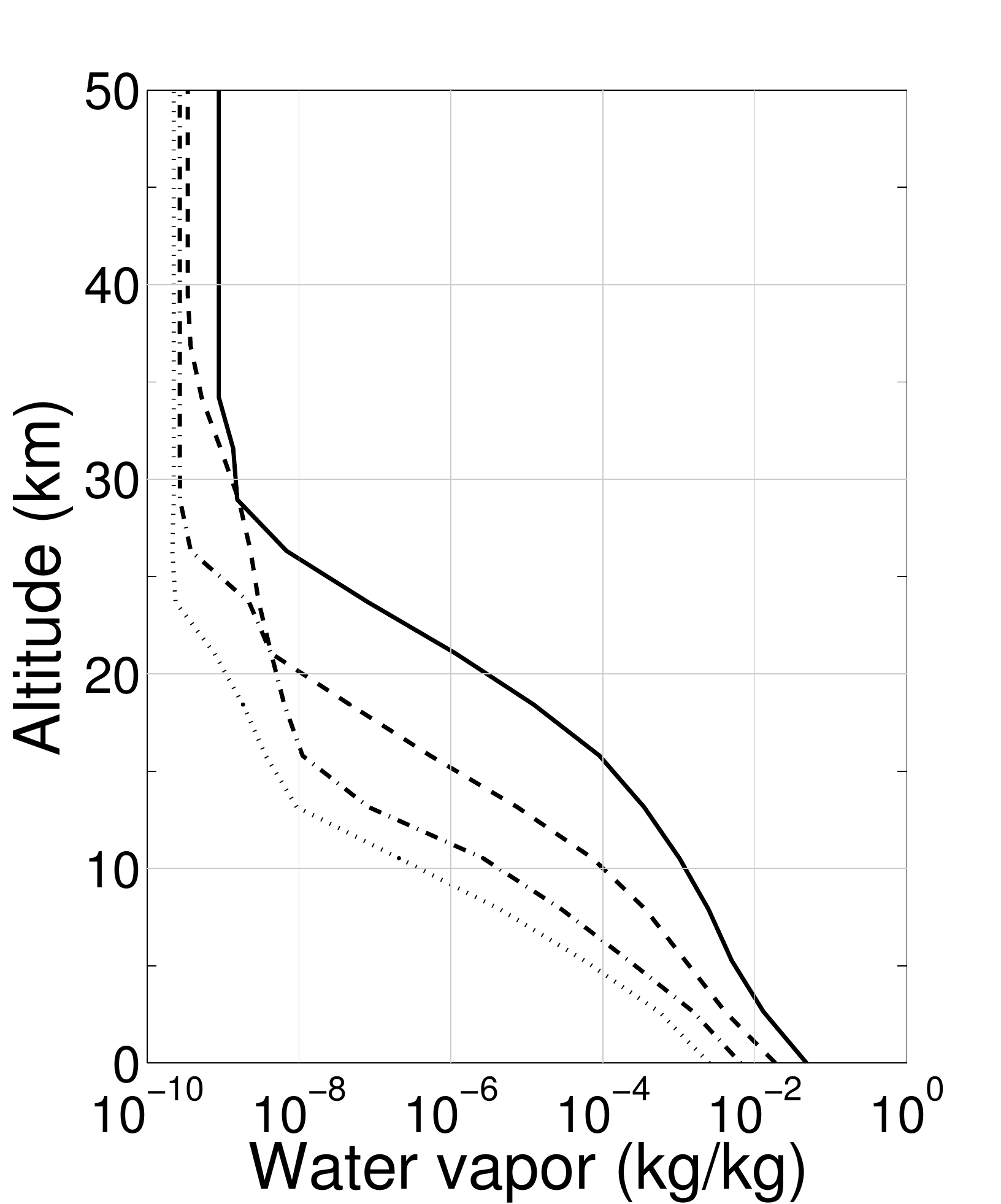}
	\includegraphics[width=5.cm]{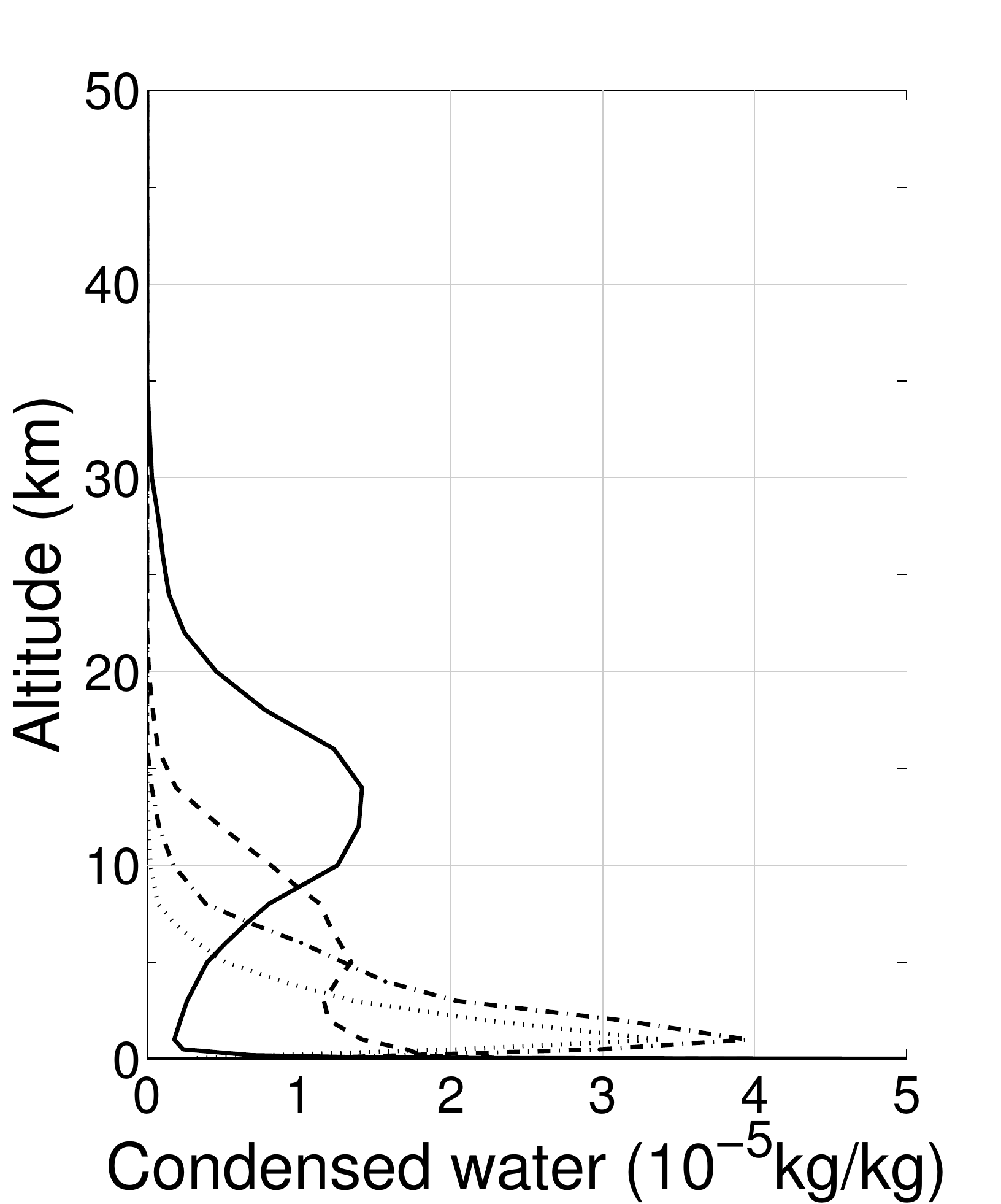}
\end{center} 
\caption{Left panel: vertical profile of global mean temperature. Middle panel: vertical profile of global mean water vapor mixing ratio (in kg/kg). Right panel: vertical profile of global mean condensed water mixing ratio (in 10$^{-5}$ kg/kg).
For the three panels, pCO$_2$ is 0.01 (dotted line), 0.1 (dashed-dotted line), 0.5 (dashed line) and 1 (solid line) bar and there is no methane.} 
\label{figure_3}
\end{figure}

\subsection{Results}
Figure \ref{figure_1} shows the global mean and zonal mean surface temperatures obtained with the GCM for present-day properties of clouds (Global mean values are gathered in table 1). Using 3.8 Ga conditions, the Earth falls into a full glaciation for the present-day pCO$_2$ (see \cite{charnay13}), but it is never fully ice-covered with at least 0.01 bar of CO$_2$. For this case with no methane, the mean surface temperature is -11.8$^\circ$C, well below the freezing point, but there is still a cold ice-free equatorial belt between -37$^\circ$N and +37$^\circ$N. With 0.1 bar of CO$_2$, the climate is temperate (global mean surface temperature around 7.4$^\circ$C without CH$_4$ and 21.3$^\circ$C with CH$_4$).
Ice-free conditions are obtained for pCO$_2$ higher than around 0.3 bar without methane or slightly higher than 0.1 bar with methane, corresponding to a mean surface temperature higher than around 23$^\circ$C. Our model has a climate sensitivity with pCO$_2$ very similar to the GCM from \cite{wolf13} going up to 0.3 bar.
With 0.5 bar of CO$_2$, the climate becomes warm (mean surface temperature around 34.7$^\circ$C without CH$_4$ and 49.9$^\circ$C with CH$_4$). With 1 bar of CO$_2$, the mean surface temperature is around 59.5$^\circ$C  without CH$_4$ and around 67.5$^\circ$C with CH$_4$.

\cite{ozak16} showed that CO$_2$ collisional line mixing reduces the atmospheric opacity between 400 and 550 cm$^{-1}$ producing a significant cooling ($\sim$10$^\circ$C) for a 1 bar CO$_2$ atmosphere for early Mars. Since we do not include collisional line mixing, our model might overestimate temperatures for high pCO$_2$. However, for the early Earth with a 1 bar CO$_2$ atmosphere, the peak of thermal emission from the surface would be around 1110 cm$^{-1}$ (compared to around 730 cm$^{-1}$ for early Mars), so quite far from the 400-550 cm$^{-1}$ spectral region. We therefore expect a weaker cooling effect by CO$_2$ collisional line mixing for the early Earth than for early Mars.

Fundamental changes to cloud properties on early Earth have been proposed as alternative solutions to the faint young Sun paradox. In particular, it has been proposed that cloud particles were larger on the early Earth because of less cloud condensation nuclei from biological sources \citep{rosing10}.
In figure \ref{figure_2}, we tested the impact of larger liquid cloud particles (17 $\mu$m versus 12 $\mu$m). Larger particles warm the planet by decreasing the planetary albedo \citep{charnay13}. This effect is particularly strong for temperate climate with a warming of around 8$^\circ$C for pCO$_2$=0.1 bar. However, it becomes weak for warmer climates. 
We also simulated the climates with a fixed amount of CCN and varying cloud radii (black line in left panel in figure 2). The mean surface temperature is almost identical to the case with 12 $\mu$m particles. 
 
The right panel in figure \ref{figure_2} shows the shortwave (albedo cooling effect) and longwave (greenhouse warming effect) radiative forcing by clouds depending on pCO$_2$ for the different particle size parametrizations. The trends are similar for all particles size parametrizations. The net cloud radiative forcing is higher for cold climates (e.g. pCO$_2$=0.01 bar) than for temperate climates (e.g. pCO$_2$=0.1 bar). This is due to the lower albedo cooling of clouds over sea ice and to a reduction of the cloud cover over the cold equatorial ocean, particularly close to the freezing line (see \cite{charnay13}). The net cloud radiative forcing increases with pCO$_2$ when pCO$_2$ exceeds 0.1 bar. The albedo cooling effect decreases because of a reduction of lower clouds (see right panel in figure \ref{figure_3}). At the same time, the greenhouse warming effect decreases. This is mostly due to the gaseous infrared emission level (the altitude where most of the thermal emission is emitted) going up with more CO$_2$, which reduces the greenhouse effect of clouds. 
The cloud albedo cooling decreases faster than the cloud greenhouse warming, which leads to a net forcing increasing with pCO$_2$. Clouds therefore produce a destabilizing feedback on the climate. For hot climates, they tend to have a small net radiative effect (their cooling effect being reduced by around 30 W/m$^2$), which depends little on the radii of cloud particles.

In response to clouds properties, our GCM predicts that hot climates can be obtained with 1 bar or less of CO$_2$ at 3.8 Ga. This required amount of CO$_2$ is significantly lower than previous 1D estimations by \cite{kasting06a}. They found that if there were no methane, around 3 bars of CO$_2$ would be required for getting a mean surface temperature of 60$^\circ$C at 3.3 Ga.  This discrepancy is due to cloud feedbacks, not simulated in the 1D model, and to a few other differences in the climate modeling.
In particular, they used the present-day land cover while we assumed no continent, and we used more recent opacity data, likely increasing the visible and infrared gaseous absorption. From a purely climate modeling point of view (i.e., ignoring long-term geochemical feedbacks), a warm early Earth with temperature of 60-70$^\circ$C is possible with sufficient CO$_2$ if there is no land or almost no land.

\section{Carbon cycle modeling}
\subsection{Model set-up and boundary conditions}

In order to quantify the carbon cycle equilibrium response to simulated climate, we used the model from \cite{lehir08a, lehir08b}. We generally assumed a classical plate tectonic regime as in \cite{sleep01} (see Fig. 1 in Appendix A), although there is debate about when a modern style of plate tectonics began \citep{stern07}. CO$_2$ reservoirs are 1) the atmosphere, 2) the surface/deep ocean, 3) the mantle, 4) carbonates on continental platforms and 5) calcite precipitation in the oceanic crust resulting from seafloor weathering. The inventory of carbon between these six reservoirs evolves through CO$_2$ outgassing by mid oceanic ridges and arc volcanoes, continental and seafloor weathering, carbonate formation/dissolution, subduction and recycling. The part of subducted carbon released by arc volcanoes depends on the recycling factor. Based on a steady state assumption, we considered an instantaneous carbon recycling in subduction zones. The model parametrizations for these processes are described in Appendix A.

We simulated the carbon cycle using the surface temperatures and precipitation-evaporation rates (P-E) from the GCM. In the absence of constraints about geography, we assumed that emerged continents are homogeneously spread over all the Earth as small continents (each grid cell has its own fraction of continents). That tends to maximize the continental weathering. We interpolated GCM temperatures and Precipitation-Evaporation (P-E) between pCO$_2$ values at each grid point.
We parametrized the seafloor weathering with a dependence on the deep ocean temperature (see \cite{coogan13} for geological evidence and \cite{krissansen-totton17} for application for the last 100 Ma), assumed to be equal to sea surface temperature at 60$^{\circ}$ latitude. Most of the weathering occuring in the upper part of the oceanic crust ($<$1km), the seafloor weathering is explicitly described between 0 and 500 m by 6 layers with an ambient temperature increasing with depth (116$^\circ$C/km).  
We assumed that the potential temperature of the mantle was around 1550$^\circ$C at 3.8 Ga compared to 1330$^\circ$C today. Formulas from \cite{flament08} (with an activation energy for the mantle viscosity of $\sim$350-400 kJ/mol) lead to a oceanic crust mass production rate $\sim$11 times higher at 3.8 Ga than today. This value is consistent with the estimation by \cite{ohta96} of a 10 times higher oceanic crust production rate during the Archean.

We assumed that the CO$_2$ flux outgassed by oceanic ridges is proportional to the oceanic crust mass production rate while the seafloor weathering, limited to the first km of the oceanic crust, is proportional to the spreading rate. For a fixed oceanic crust mass production rate, a thicker crust has no impact on the outgassed CO$_2$ flux but reduces the seafloor weathering in our model, leading to an enhancement of pCO$_2$. 
It is generally accepted that the oceanic crust was thicker during the Archean ($\sim$18 km versus 6 km in average) \citep{bickle86, ohta96, flament08}.
If we assume that the early oceanic crust was 3 times thicker at the mid-oceanic ridge, the spreading rate (SR) must have been 3.8 times higher than today, according to our estimation of the oceanic crust mass production rate. Since the oceanic crust thickness and the associated spreading rate during the Archean are highly uncertain \citep{bickle86}, we therefore tested cases with the present-day oceanic crust thickness (6 km) and SR=11.3.

For a warmer mantle and a higher spreading rate on the early Earth, the younger oceanic lithosphere was warmer and melted more easily when subducted. 
If arc volcanoes, associated with subduction zones, provide a quick return of materials to the surface, the direct melting of oceanic crust may increase the amount of fluid released, suggesting a recycling of suducted carbon higher than today. According to this assumption, the decarbonation efficiency may exceed Cenozoic values used as reference \citep{hoareau15}.
We thus explored the effect of the recycling (R) using the Cenozoic value as reference (R=0.1, corresponding to a recycling efficiency of 10$\%$) and values 3 times higher (R=0.3, a recycling efficiency of 30$\%$). 

The fraction of emerged land during the Hadean/Archean is still debated \citep{roberts15}. We made the assumption that this fraction was low ($<$30$\%$ of the present-day fraction) during the early Archean as suggested by some geochemical data (e.g. \citep{dhuime12}), some evidence for a more voluminous ocean during the Archean \citep{pope12} and some models (e.g. \citep{flament08}). This hypothesis allows us to use the outputs from the 3D land-free simulations for the carbon cycle model.
We explored the effect of the emerged continental fraction considering four cases: 
\begin{itemize}
\item 1) an aquaplanet (no land)
\item 2) a very low emerged land fraction (1$\%$ of the present-day land fraction)
\item 3) a moderate continental fraction (12$\%$ of the present-day land fraction)
\item 4) a relatively high continental fraction (26$\%$ of the present-day fraction)
\end{itemize}

Although there is clear evidence of emerged land at 3.7-3.8 Ga \citep{nutman15}, an aquaplanet model is an  idealized case, which is very useful for understanding the effect of seafloor weathering.

\subsection{Results and discussion}

We ran the climate-carbon model for the four land fractions (0, 0.01, 0.12 and 0.26 times present-day fraction), with a recycling factor R equal to 1 (present-day) or 3, with or without 2 mbars of atmospheric methane and with the present-day oceanic crust thickness (H=6 km, SR=11.3) or a 3 times thicker oceanic crust (H=18 km, SR=3.8). The equilibrium points span a temperature range from 8.5$^\circ$C to 31.5$^\circ$C and a pCO$_2$ range from 0.12 bar to 0.36 bar (see figure \ref{figure_4} and table 1). Ice-free or partly ice-free oceans are obtained in any cases.

\subsubsection{Cases with no land}
\begin{figure}[!ht]
\begin{center} 
	\noindent\includegraphics[width=10cm]{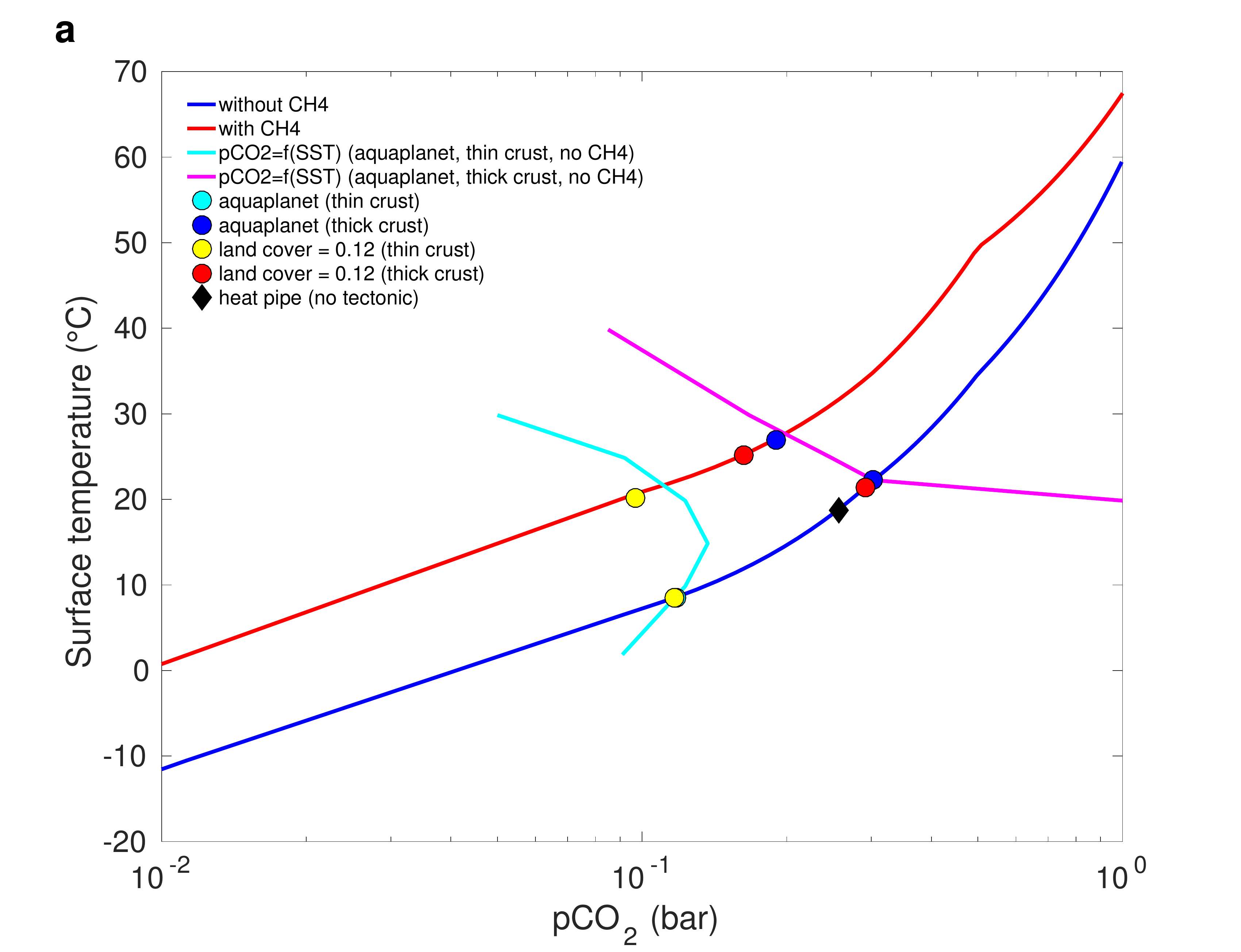}
	\noindent\includegraphics[width=10cm]{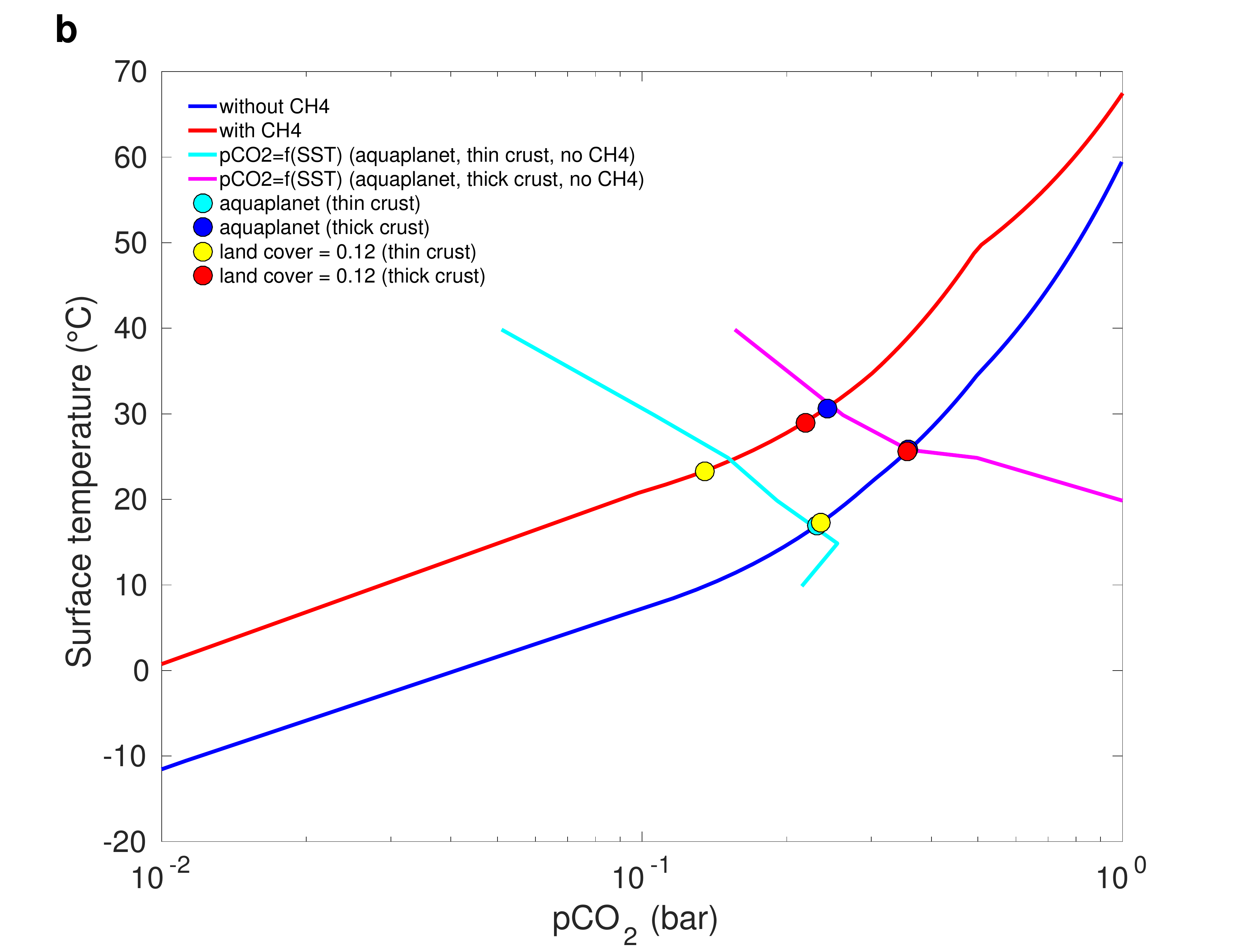}
\end{center} 
\caption{Global mean surface temperature as figure 1 but including the equilibrium points of the carbon model assuming a land fraction of 0.01 (light blue and blue circles) and 0.12 (yellow and red circles) compared to the present-day land fraction. Panel a) shows equilibrium points for the present-day recycling (R=0.1). Light blue and yellow circles are for the present-day oceanic crust thickness while blue and red circles for a 3 times thicker oceanic crust.
The black diamond corresponds to the heat pipe scenario with no plate tectonic on an aquaplanet.
Panel b) is similar to panel a) but with a 3 times higher recycling (R=0.3). 
The light blue and the purple curves are the intrinsic response of the carbon cycle model. They are obtained by changing the surface temperature from 10$^\circ$C to 40$^\circ$C with no feedback of pCO$_2$ on the temperature, with no land, for the present-day oceanic crust (light blue) and for a 3 times thicker oceanic crust (purple).} 
\label{figure_4}
\end{figure}

If there is no emerged land, the carbon cycle is controlled by the seafloor weathering and volcanic/mid-oceanic ridge outgassing. At equilibrium, the seafloor weathering is simply given by $F_{sfw}=\frac{F_{mor}+F_{deep}}{1-R}$ ($F_{mor}$ and $F_{deep}$ are the CO$_2$ fluxes from midocean ridges and mantle, respectively, see Appendix A) and depends on $R$. Because the carbon species partitioning between ocean and atmosphere is controlled by temperature, sea surface temperature (SST) implicitly determines the seawater pH and pCO$_2$ for a fixed value of seafloor weathering rate. 
Light blue and purple curves in figure \ref{figure_4} reveal the CO$_2$ partial pressure for a given SST with our model. They are computed assuming no CO$_2$ climate feedback and thus represent the intrinsic response of the carbon model for Earth's climate in the absence of continents.
The light blue curves (6 km thick crust) reveal a maximum pCO$_2$ of 0.14 bar for R=0.1 (left panel) and 0.26 bar for R=0.3 (right panel). The CO$_2$ solubility in the ocean decreases with temperature. At temperatures lower than 15$^\circ$C, this feedback dominates and pCO$_2$ increases with temperature. When the temperature becomes higher than 15$^\circ$C, the temperature dependance of the seafloor weathering dominates and pCO$_2$ decreases with temperature. The combination of both effects (solubility and seafloor weathering) leads to an upper limit for pCO$_2$ for the thin oceanic crust case.
The reduced seafloor accretion rate associated to a thicker oceanic crust scenario combined with cold water strongly limits carbon consumption by seafloor weathering, hence the purple curves (18 km thick crust) do not show a maximum pCO$_2$ below 1 bar.
For these cases without land, the equilibrium point of the climate-carbon model corresponds to the crossing between the intrinsic response of the carbon model (i.e. the light blue and purple curves) and the GCM curves showing the global mean surface temperature as a function of pCO$_2$ (i.e. the blue and red curves). For instance, the equilibrium point for a methane-free atmosphere with R=0.1 and a thick crust corresponds to a mean SST of 22.3$^\circ$C and pCO$_2$=304 mbar.

\subsubsection{Effects of land cover}
According to our model, land cover has only a limited impact on the carbon cycle and the global climate for the early Earth (temperature change of $\sim$1$^\circ$C between the four land distributions).
Indeed, while the carbon cycle was mostly controlled by continental weathering during the Proterozoic and the Phanerozoic (e.g. \cite{krissansen-totton17}), it was mostly controlled by seafloor weathering on the early Earth. That difference is mainly due to the higher spreading rate (high accretion rate provides a sufficient quantity of weatherable materials which enhances the seafloor weathering) and to the assumed lower land fraction on the early Earth.
Under such circumstances, our assumption about the land distribution (i.e. small continents distributed homogeneously) is therefore of minor importance for the results.
We also notice that the feedback produced by land does not have the same sign with a 6 or 18 km thick oceanic crust. With a 6 km oceanic crust, the seafloor weathering is slightly stronger than the continental weathering per unit of surface. In that case, the global mean surface temperature and pCO$_2$ are higher with a higher land fraction. In contrast, with a 18 km thick oceanic crust, the continental weathering is stronger than the seafloor weathering per unit of surface (as today) and a higher land fraction leads to lower pCO$_2$.
The coldest case with our assumptions (mean SST of 8.5$^\circ$C) is thus obtained with no land, a methane-free atmosphere, R=0.1 and a thin crust. The warmest case (mean SST of 31.5$^\circ$C) is obtained with no land, a methane-rich atmosphere, R=0.3 and a thick oceanic crust, as shown by the blue point in Figure 4b.

\subsubsection{Effects of methane, oceanic crust thickness and recycling rate}
A thicker oceanic crust or a higher recycling efficiency always leads to an increase of pCO$_2$ and so to an increase of temperature. In contrast, adding methane in the atmosphere reduces pCO$_2$ but still increases the global mean surface temperature. 
If we use as a reference, the case with a moderate land cover, a 6 km thick crust, no methane and R=0.1 (mean surface temperature=8.5$^\circ$C), adding 2 mbar of methane produces a 11.7$^\circ$C warming, a 18 km crust produces a 12.9$^\circ$C warming, and a 3 times higher recycling rate produces a 8.8$^\circ$C warming. These three parameters therefore produce significant temperature and pCO$_2$ changes. In particular, the thicker oceanic crust is generally expected for the early Earth, with an effect on pCO$_2$. A thicker oceanic crust would have had a large impact, allowing warmer climate for the early Earth than today.
The combination of a thicker crust, a high recycling rate, and the presence of methane leads to global mean surface temperatures around 30$^\circ$C for the three land distributions. That temperature change ($\sim$20$^\circ$C) is much lower than the sum of individual warming ($\sim$33$^\circ$C), mostly because of the non-linear dependence of the seafloor weathering with the deep ocean temperature.

\subsubsection{Absence of plate tectonic: the heat-pipe scenario}
The onset time of plate tectonic is unknown but most geological and isotopic data suggest that subduction started during the Hadean or early Archean \citep{sleep07, tang16}. \cite{moore13} suggested that before the onset of plate tectonic, Earth was in a heat-pipe regime in which volcanism dominated surface heat transport. We tested the consequences of this possible early regime for the carbon cycle with our model. We considered that all the heat flux is released by volcanism, that the spreading rate is null and that submarine basalt flows are weathered as seafloor basalts. In our simulation of the heat-pipe Earth, we assumed a very low land fraction (1$\%$ of the present-day land cover). The volcanic outgassing becomes around 4 times as high as with the plate tectonic regime, or 10 times as high as the present volcanic outgassing. At equilibrium, pCO$_2$ is 0.26 bar and the mean surface temperature is around 18.8$^\circ$C. A heat-pipe regime would thus have led to a temperate early Earth.

\subsubsection{Discussion}
Our climate-carbon model never reaches warm climates for the early Earth because of the strong temperature feedback by the seafloor weathering. It predicts global mean surface temperatures around 8.5-30.5$^\circ$C, with pCO$_2$ around 0.1-0.36 bar. This temperature range is compatible with the most recent measurements by \cite{hren09} and \cite{blake10}, and also with the evolutionary model from \cite{boussau08} for a mesophilic last universal common ancestor (LUCA). Thermophily near the roots of the tree of life may reflect local warm environments (like hydrothermal vents) or survival after hot climates produced just after large impacts \citep{boussau08, abramov09}. 
The pCO$_2$ range of 0.1-0.35 bar is generally higher than the upper limits initially derived from the stability of different minerals from Archean samples \citep{rye95, sheldon06, rosing10, driese11}. These constraints globally suggest a pCO$_2$ lower than 30 mbars. However, they are not all compatible together and were essentially obtained for the late Archean. Recent work by \cite{kanzaki15} gives an upper limit for pCO$_2$ of 0.14-0.7 bar at 2.77-2.46 Ga, compatible with our results.

\begin{table}
\begin{footnotesize}
\begin{tabular}{|m{0.9cm}|m{1.3cm}|m{0.5cm}|m{1.0cm}|m{0.5cm}|m{0.5cm}||m{0.8cm}|m{0.8cm}|m{0.5cm}|m{0.6cm}|m{0.5cm}|m{0.6cm}|m{0.7cm}|m{0.7cm}|}
\hline
  \multicolumn{6}{|c||}{Boundary conditions} & \multicolumn{8}{c|}{Results} \\
\hline

Land fraction	& Oceanic crust (km)	&	SR		&	Recyling  	&	CH$_4$	& rdeg 	&pCO$_2$ (mbar) 	&	Tsurf ($^\circ$C) 	&pH 		& DIC 	& F$_{mor}$	& F$_{arc}$ 	& F$_{con}$	& F$_{sfw}$	\\ \hline
1  			& 6					&	1 		&	0.1 		&	no    	&1 		&0.285 			&	14.9				&8.4	 	& 2.4	& 1.6		& 7.7		& -10.5		& -1.1		\\ \hline
0.0  			& 6					&	11.3 	&	0.1 		&	no    	&2.5 	&118 			&	8.5				&6.70	& 32.2 	& 18			& 11.2		& 0.0		& -29.2		\\		
0.0  			& 6					&	11.3 	&	0.3 		&	no    	&2.5 	&225			&	16.5				&6.60	& 39.8 	& 18			& 19.5		& 0.0		& -37.5		\\
0.0  			& 18					&	3.8		&	0.1 		&	no    	&2.5 	&304			&	22.3				&6.56	& 48.9	& 18			& 11.2		& 0.0		& -29.2		\\		
0.0  			& 18					&	3.8	 	&	0.1 		&	yes    	&2.5 	&190 			&	27.0				&6.66	& 35.4	& 18			& 11.2		& 0.0		& -29.2		\\	
0.0  			& 18					&	3.8	 	&	0.3 		&	no    	&2.5 	&359 			&	25.8				&6.53	& 52.4	& 18			& 19.5 		& 0.0		& -37.5		\\
0.0  			& 18					&	3.8		&	0.3 		&	yes    	&2.5 	&243 			&	30.6				&6.61	& 39.6	& 18			& 19.5		& 0.0		& -37.5		\\ \hline	
0.01  		& 6					&	11.3 	&	0.1 		&	no    	&2.5 	&118 			&	8.5				&6.70	& 32.21 	& 18			& 11.2		& -0.05		& -29.2		\\		
0.01  		& 6					&	11.3 	&	0.3 		&	no    	&2.5 	&231			&	16.9				&6.65	& 45.10 	& 18			&			&			&			\\
0.01  		& 18					&	3.8		&	0.1 		&	no    	&2.5 	&303			&	22.3				&6.56	& 49.51	& 18			&			&			&			\\		
0.01  		& 18					&	3.8	 	&	0.1 		&	yes    	&2.5 	&186 			&	26.8				&6.67	& 35.39	& 18			&			&			&			\\	
0.01  		& 18					&	3.8	 	&	0.3 		&	no    	&2.5 	&358 			&	25.8				&6.53	& 52.73	& 18			&			&			&			\\
0.01  		& 18					&	3.8		&	0.3 		&	yes    	&2.5 	&240 			&	30.5				&6.62	& 39.64	& 18			&			&			&			\\ \hline
0.12  		& 6					&	11.3 	&	0.1 		&	no    	&2.5 	&117 			&	8.5				&6.70 	& 32.06	& 18			& 11.6		& -0.4		& -29.2		\\		
0.12  		& 6					&	11.3 	&	0.1 		&	yes    	&2.5 	&97 				&	20.2				&6.80	& 25.52	& 18			&			&			&			\\
0.12  		& 6					&	11.3 	&	0.3 		&	no    	&2.5 	&235 			&	17.3				&6.66	& 46.69	& 18			& 23.7		& -3.1		& -38.7		\\	
0.12  		& 6					&	11.3 	&	0.3 		&	yes    	&2.5 	&135 			&	23.3				&6.75	& 31.47	& 18			&			&			&			\\
0.12  		& 18					&	3.8 		&	0.1 		&	no    	&2.5 	&292 			&	21.4				&6.59	& 50.13	& 18			& 12.6		& -4.1		& -26.5		\\		
0.12  		& 18					&	3.8	 	&	0.1 		&	yes    	&2.5 	&163 			&	25.2				&6.71	& 34.38	& 18			&			&			&			\\	
0.12  		& 18					&	3.8 		&	0.3 		&	no    	&2.5 	&356 			&	25.6				&6.57	& 56.31	& 18			&			&			&			\\
0.12  		& 18					&	3.8	 	&	0.3 		&	yes    	&2.5 	&219 			&	29.0				&6.69	& 41.89	& 18			&			&			&			\\ \hline
0.26  		& 6					&	11.3 	&	0.1 		&	no    	&2.5 	&121 			&	8.8				&6.70	& 32.50	& 18			& 12.1		& -0.64		& -29.5		\\
0.26	 		& 6					&	11.3 	&	0.3 		&	no    	&2.5 	&245 			&	18.0				&6.66	& 48.47	& 18			&			&			&			\\ \hline			
0.12  		& 6					&	0		&	0              &	no    	&10 		&257 			&	18.8				&6.60	& 45.75	& 31.9		& 1.1 		& -0.9		& -32.1		\\
\hline
\end{tabular}
\end{footnotesize}
\label{table1}
\caption{Equilibrium states from the carbon model coupled to the GCM. Columns from left to right are: the land fraction, the oceanic crust thickness (in km), the spreading rate, the recycling rate, the presence (2 mbar) or absence of methane, the outgassing rate, pCO$_2$ (in mbar), the mean surface temperature (in $^\circ C$), the mean oceanic pH, the mean dissolved inorganic carbon (in mol/m$^3$), the CO$_2$ flux outgassed from mid-oceanic ridges and from arc volcanoes, and the CO$_2$ sink by continental weathering and seafloor weathering (all fluxes in 1e12 mol/yrs). The first case, with land fraction equal to 1, is the present-day Earth. The last case, with SR=0, is the heat-pipe scenario with no plate tectonics.}
\end{table}

\section{Climate and carbon cycle during the Late Heavy Bombardment}
\subsection{Impact modeling}
The Earth was likely struck by numerous of impactors during the Late Heavy Bombardment (LHB), which would have occurred between around 4.1 and 3.8 Ga \citep{gomes05, bottke17}. 
These impactors may have been a sink of CO$_2$ by producing large amounts of ejecta, which can be easily weathered \citep{sleep01}. They may also have been a source of CO$_2$ by vaporizing carbonates from the oceanic crust. For instance, Chixulub impact may have outgassed 350-3500 Gt of CO$_2$ \citep{pierazzo98}, or the equivalent of 0.06-0.6 mbar of CO$_2$.
We investigated the long term effects of impacts on the carbon cycle to understand their consequences for the mean climate during the LHB, in particular if they had a global net warming or cooling effect.

We assumed that the frequency of impact follows a scaling law given in \cite{anbar01} and \cite{sleep01}:

\begin{eqnarray}
\it f(\geq m) = f_0 \times \left(\frac{m}{m_{max}}\right)^{-b}
\end{eqnarray}
\\
where f is the frequency of impactors having a mass higher than $m$.
For the LHB, we chose $b$ = 0.7 \citep{anbar01}. We considered that the maximal mass was $m_{max}=10^{19}$ kg (i.e. diameter around 190 km). This upper limit is constrained by the largest crater observed on the Moon. We neglected impacts with a mass lower than 10$^{13}$kg (i.e. diameter lower than around 1.8 km), which produce little ejecta because of the strong deceleration by the ocean.
We fixed $f_0$ = 2.9$\times 10^{-8}$ impacts/year in order to have a cumulative impactor mass equal to 2$\times 10^{20}$ kg for the 300 millions of years of the LHB. That corresponds to an impact flux of 6.7$\times 10^{11}$ kg/year.

As \cite{sleep01} and \cite{zahnle02}, we computed the weathering of ejecta, which provides a source of cations, precipitating as carbonates in the ocean. 
We assumed that the mass distribution of ejecta is given by $\it N(\geq m) \propto m^{-\gamma}$, where $\gamma$=0.87 (derived from scaling laws of \cite{collins05}). \cite{zahnle02} used a similar law with $\gamma$=0.9-0.95. We considered that the diameter of each particle decreases of 1 mm in 167 kyrs \citep{crovisier87} and that cations are present at 0.005 mol/g. With the oceanic crust lifetime for the early Earth, ejecta with a diameter larger than 100 m contribute little to the weathering compared to smaller ejecta.

We included CO$_2$ delivered by the impactors as done by \cite{deniem12}. We assumed that the mass fraction of CO$_2$ is 1$\%$  in asteroids and 20$\%$ in comets. For comparison, Rosetta measurements of Comet 67P / Churyumov-Gerasimenko indicate a mass ratio of CO$_2$ to H$_2$O around 40$\%$ \citep{marty16} for a mass fraction of H$_2$O to dust around 20$\%$, implying a mass fraction of CO$_2$ around 8$\%$. If all impactors were comets, the LHB would  have provided the equivalent of 3.8 bar of CO$_2$ (taking into account the escaping mass of impactors). The fraction of comets during the LHB is not known and \cite{bottke12} suggest that comets were a minor player. We did simulations using fractions of comets of 0 and 100$\%$.
We also included the CO$_2$ outgassed by oceanic carbonates and lithosphere melting after impact. We considered that the oceanic crust is covered by a 40 m layer of carbonate, which outgassed CO$_2$ when shocked at a high pressure (above 60 GPa). We assumed that the lithosphere below the carbonate layer (considered as pure calcite) contains 100 ppmv of CO$_2$ and outgasses all CO$_2$ when melted.
The complete assumptions for the ejecta weathering and CO$_2$ outgassing are described in Appendix B. 

Figure \ref{figure_5} shows the amount of CO$_2$ outgassed from impactor/crust and consumed by ejecta as a function of impactor mass for asteroids (density=3000 kg/m$^2$ and initial velocity=21 km/s) and comets (density=1000 kg/m$^2$ and initial velocity=30 km/s). The solid red lines show the maximal amount of CO$_2$ that can be consumed by ejecta. The solid red line show the amount of CO$_2$ consumed by ejecta in 15 Ma (the lifetime of the oceanic crust in our model). With our size distribution for ejecta, most of ejecta are weathered during the lifetime of the oceanic crust. In all cases, impacts produce a net sink of CO$_2$ by ejecta weathering.

\subsection{Results and discussion for the LHB}
To simulate the effect of impacts during the LHB, we computed sequences of 10 millions of years with stochastic impacts following the statistical distribution described above. For simplicity, we assumed that impacts never overlap and that they all fall into the ocean. This assumption is valid because we consider small continental fractions and because the total surface of transient craters (diameter $D_{tc}$ in Appendix B) accumulated in 10 million years only represents 0.6 $\%$ of the Earth's surface.
These sequences provide a time-dependent CO$_2$ outgassing rate and alkaline source, which are used in the carbon cycle model to computed the climate and pCO$_2$ evolution.
The top panel in figure \ref{figure_6} shows the cumulated CO$_2$ outgassed by impacts and consumed by ejecta during a typical 10 million year sequence with stochastic impacts. In this sequence, one 143 km impactor strikes the Earth at 7 Ma and dominates CO$_2$ fluxes. 
On average, the weathering of ejecta produces a cations flux of 16 $\times 10^{12}$ mol/yr (compared to 300 $\times 10^{12}$ mol/yr in \cite{sleep01}), around half the CO$_2$ flux from mid oceanic ridges and arc volcanoes (around 30$\times$10$^{12}$mol/year). The vaporization of carbonates produces a CO$_2$ flux of 0.05$\times$10$^{12}$mol/year, negligible on the long term compared to the other fluxes.
The bottom panel in figure \ref{figure_6} shows the evolution of pCO$_2$ during this sequence for the 0.01 land fraction, no methane, R=0.1, a thin or thick oceanic crust and with asteroids or comets.
The giant impact at 7 Ma produces a strong decrease of pCO$_2$ by ejecta weathering triggering a full glaciation for thin oceanic crust cases and a very cold climate with a global mean surface temperature of -10$^\circ$C for the thick oceanic crust case. Before this impact, the carbon cycle is stabilized with pCO$_2$ around 50 mbar for the thin crust and around 250 mbar for the thick crust.
As in \cite{sleep01} and \cite{zahnle02}, we thus find a strong cooling by ejecta weathering leading to cold climates, but not always completely ice covered. In particular, cases with a thick crust, methane and R=0.3 should remain temperate or at least partially ice-free most of the time during the LHB.

To better understand the role of each parameter, we derived a simple analytical expression of pCO$_2$. If we assume a case with no land, a recycling R$_0$ and an oceanic crust thickness H$_0$, the equilibrium of the carbon cycle can be expressed for a reference state with no impact as:

\begin{eqnarray}
\it F_{mor}^0+F_{arc}^0+F_{sfw}^0=0
\end{eqnarray}
\\
where $F_{mor}^0$, $F_{arc}^0$ and $F_{sfw}^0$ are the CO$_2$ fluxes from mid-oceanic ridges, arc volcanoes and seafloor weathering from the (crust+mantle) to the (ocean+atmosphere).
For a new state with other values of R and spreading rate and with impacts, the equilibrium is given by:

\begin{eqnarray}
\it F_{mor}+F_{arc}+F_{sfw}+F_{ej}=0
\end{eqnarray}
\\
with $F_{mor}=F_{mor}^0$ and $F_{arc}=F_{deep} -R\times (F_{sfw}+F_{ej})$ (see appendix A).
We make the assumption that the seafloor weathering does not depend on oceanic temperature, what is valid for cold climates. By using the simple expression: $F_{sfw}=F_{sfw}^0\left(\frac{pCO_{2}}{pCO_{2}^0} \right)^{\alpha}\left(\frac{SR}{SR_0} \right)$  with $\alpha$=0.5 \citep{haqq-misra16}, we obtain an expression for pCO$_2$:

\begin{eqnarray}
\it pCO_2=pCO_2^0 \times \left(\frac{SR_0}{SR} \right)^{1/\alpha}\left(\frac{1-R_0}{1-R} \right)^{1/\alpha}\left(1-\frac{(1-R)\times F_{ej}}{F_{mor} + F_{deep}} \right)^{1/\alpha}
\end{eqnarray}

The term $\frac{(1-R)\times F_{ej}}{F_{mor} + F_{deep}}$ simply corresponds to the ratio of the amount of CO$_2$ lost in the mantle (due to ejecta weathering) by the amount of CO$_2$ outgassed directly from the mantle (e.g. without taking into account recycling). When this ratio is higher than 1, there is a net sink of CO$_2$ toward the mantle and the Earth necessary falls into a full snowball Earth. That occurs for $F_{ej}\leqslant$ -29$\times$12 mol/yr for R=0.1 and $F_{ej}\leqslant$ -37$\times$12 mol/yr for R= 0.3 with our model values.
With our impact distribution, the ejecta cations flux (16 $\times$ 10$^{12}$ mol/yr) is below these limits, and the Earth should have avoided a permanent full glaciation during the LHB. 
Because of the large uncertainties on the carbon cycle parameters and the duration and intensity of the LHB, it is not possible to be definitive on that point.
We conclude that the LHB likely had a strong impact on the carbon cycle and climate with potentially snowball Earth events after large impacts. 

Large impactors were also frequent during all the Archean \citep{bottke12, johnson12, bottke17} but much less than during the LHB. According to the dynamical model from \cite{bottke12}, the impact frequency was an order of magnitude lower at 3.7 Ga than at the peak of the LHB. Therefore, impacts probably had only a limited effect on the long term carbon cycle and climate during most of the Archean. However, our LHB impact sequence reveals that individual giant impacts could trigger episodic glaciations. \cite{johnson12} evaluated the diameters of ancient large impactors from spherule layers during the Archean and the Proterozoic. A few of them with diameters higher than around 50 km could have been large enough to trigger glaciations.

The existence of the Late Heavy Bombardment is still debated by some authors. For instance, \cite{Boehnke16} suggest that the LHB could be an artifact of monotonically decreasing impact flux combined with episodic crust formation. If it did not occur, the impact flux at the end of the Hadean would have been too small to affect the climate on the long term, but strong cooling after episodic large impacts may still have been possible.

\begin{figure}[h!]
\begin{center} 
	\noindent\includegraphics[width=10cm]{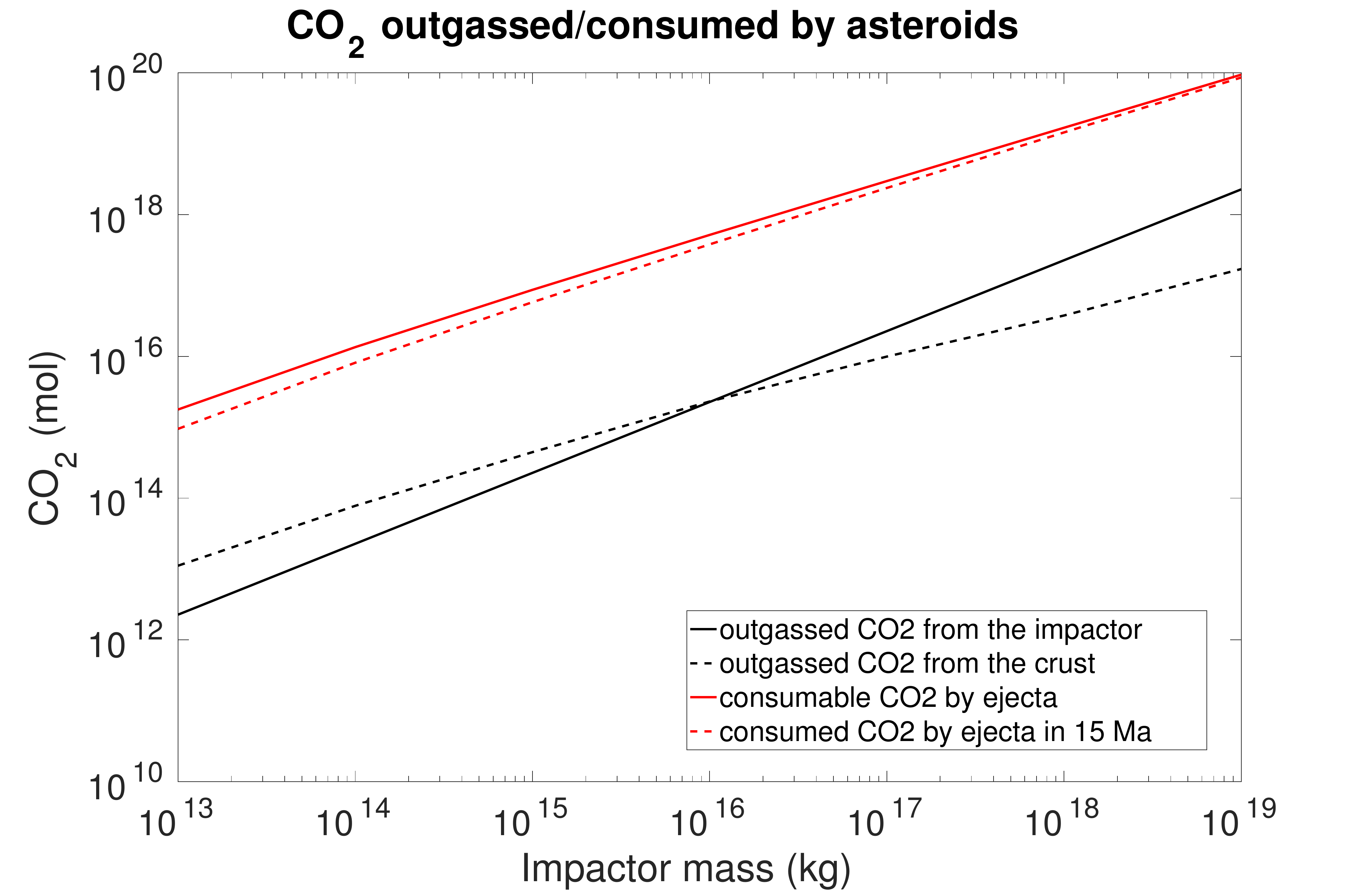}
	\noindent\includegraphics[width=10cm]{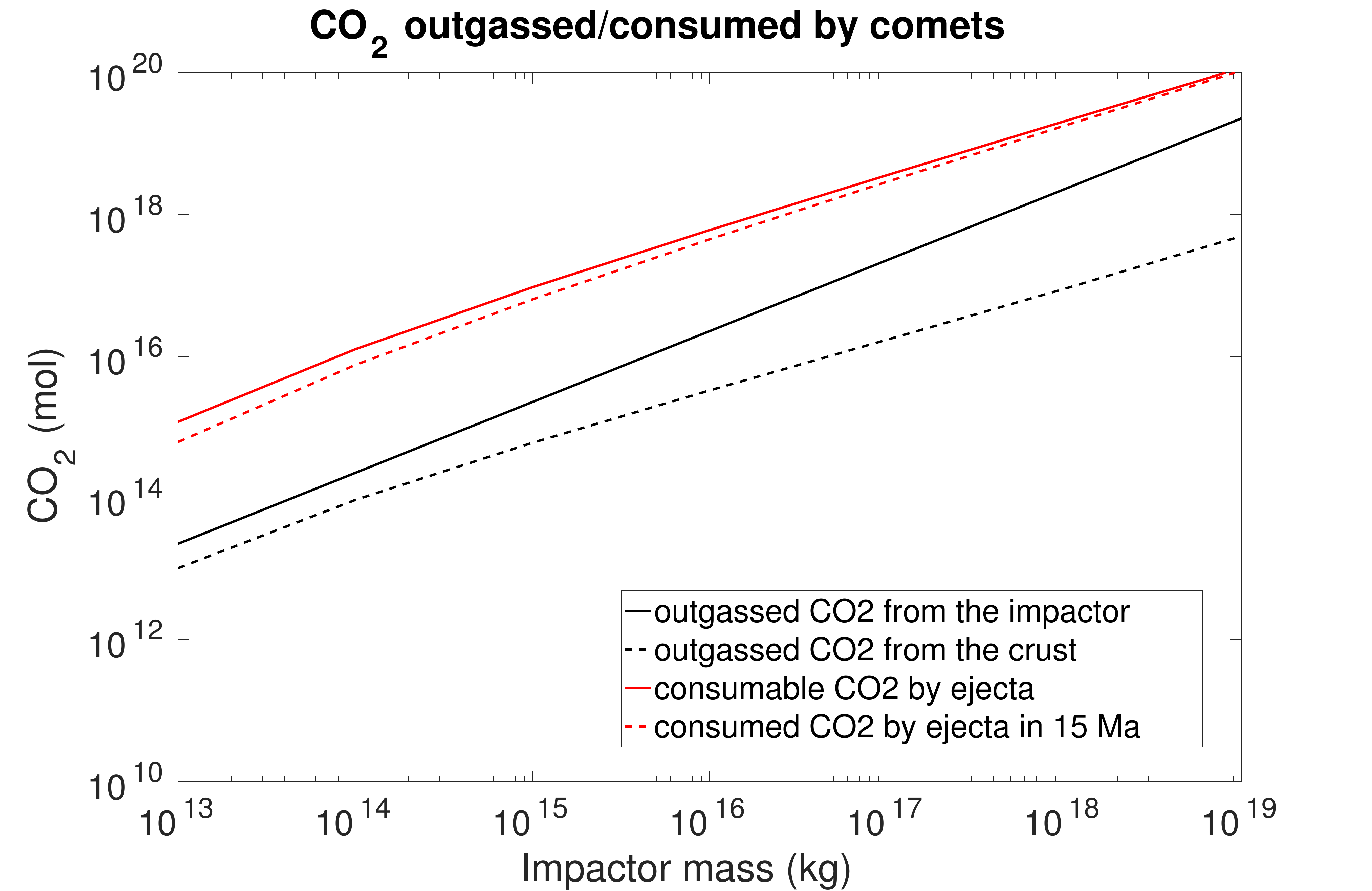}
\end{center} 
\caption{Effect of an individual impact on atmospheric CO$_2$.
Amount of CO$_2$ outgassed from impactor (black solid lines) and from the crust (black dashed lines) or consumed by ejecta weathering as a function of impactor mass for asteroids (top panel) and comets (bottom panel). The red solid line is the maximal amount of CO$_2$ that can be consumed by ejecta weathering, while the red dashed line is the amount of CO2$_2$ consumed during 15 Ma (approximately the lifetime of the Archean oceanic crust).} 
\label{figure_5}
\end{figure} 

\begin{figure}[h!]
\begin{center} 
	\noindent\includegraphics[width=10cm]{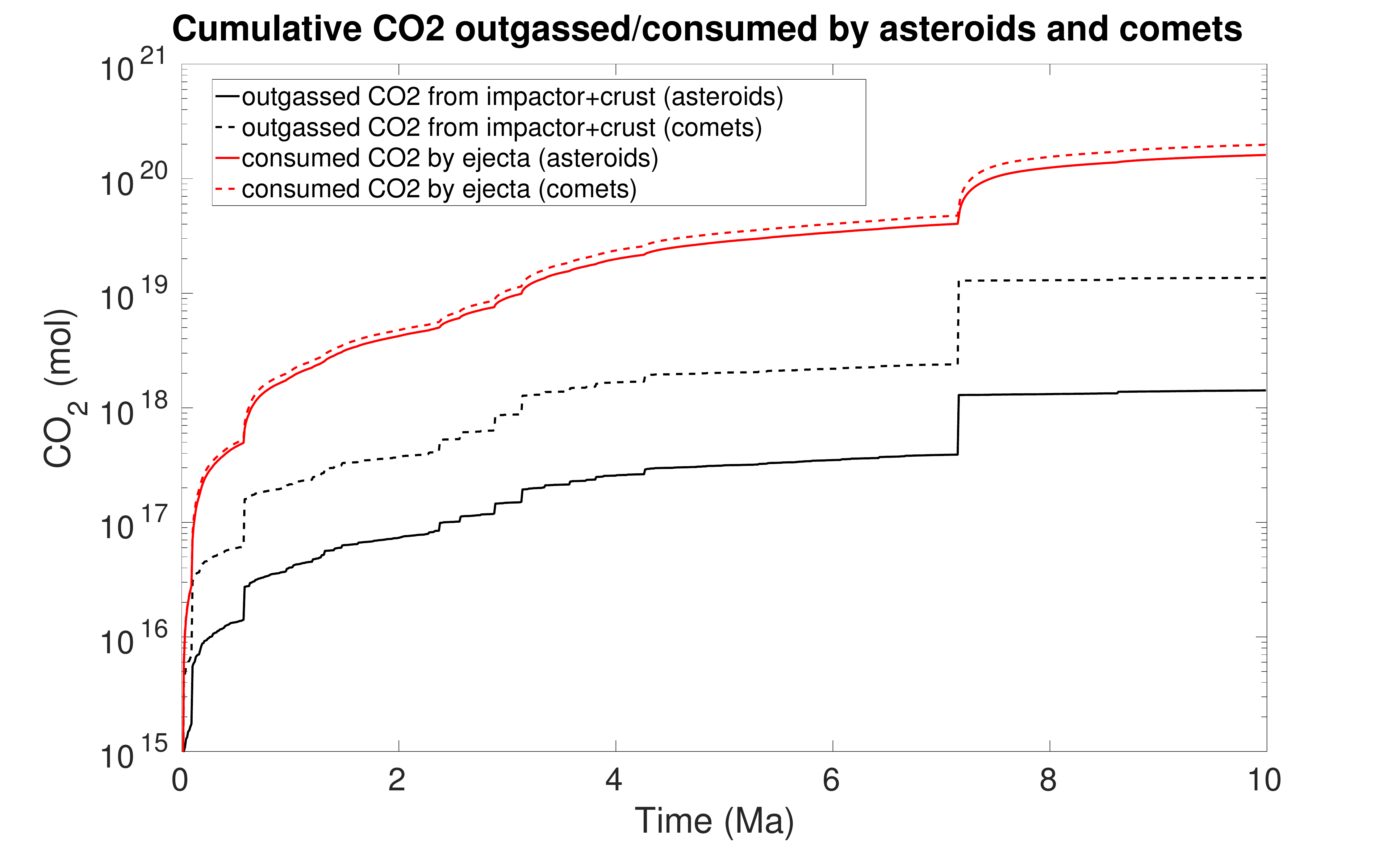}
	\noindent\includegraphics[width=10cm]{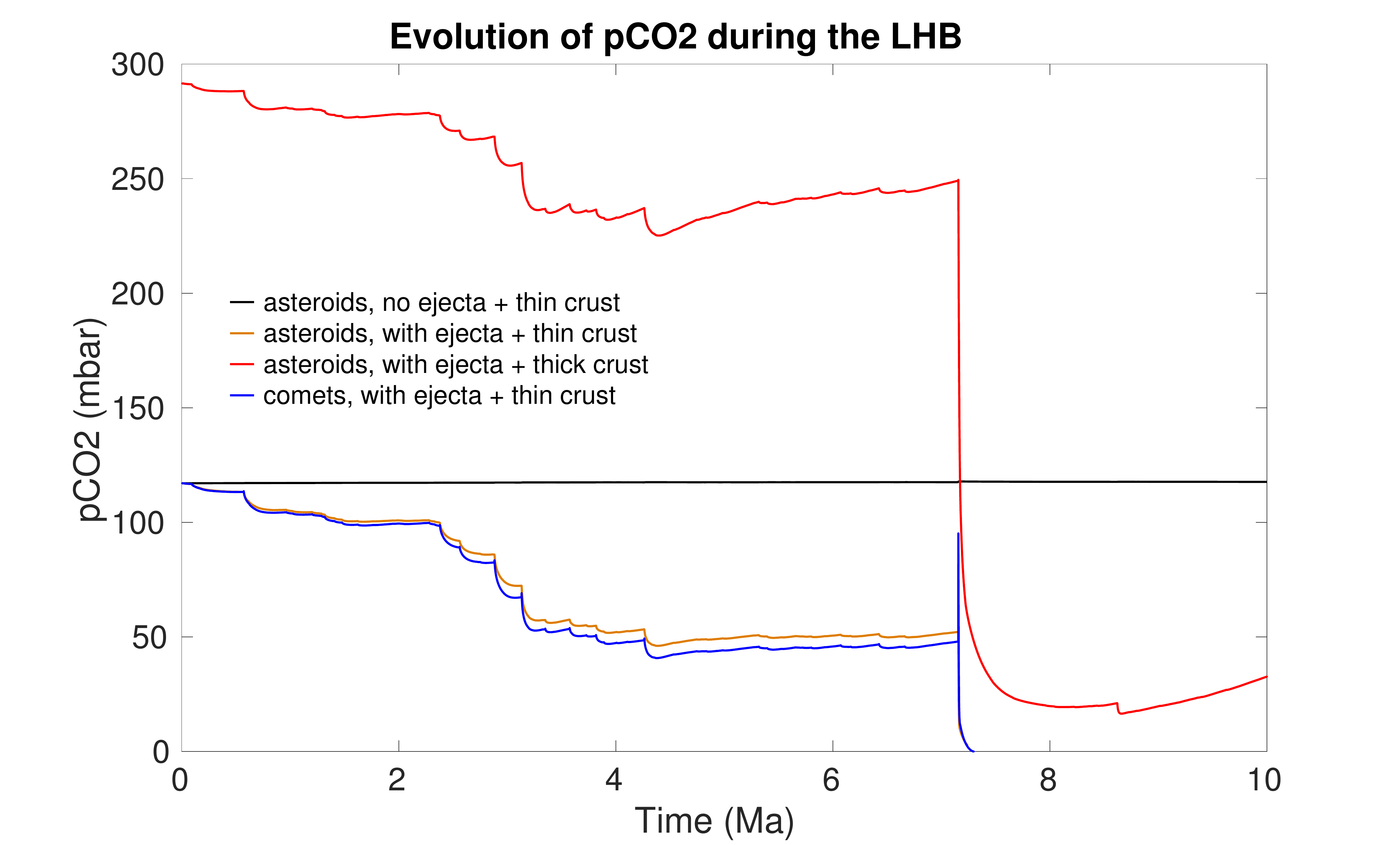}
\end{center} 
\caption{Effect of impacts on pCO$_2$.
Top panel: impact scenario expressed in cumulative mol of CO$_2$ outgassed by impacts (black line) and consumed by ejecta (red line) with asteroids (solid lines) or comets (dotted lines).
Bottom panel: evolution of pCO$_2$ for the impact scenario, assuming a low land fraction, R=0.1 and no methane. 
Yellow line is a case with a thin oceanic crust, asteroids and ejecta weathering.
Black line is as yellow line but without ejecta weathering.
Blue line is as yellow line but with comets.
Red line is as yellow line but with a thick oceanic crust.} 
\label{figure_6}
\end{figure} 

\section{Conclusions}
Our 3D climate simulations (taken in isolation) reveal that warm climates, with mean surface temperatures around 60$^\circ$C, could have been obtained during the late Hadean/early Archean with around 1 bar of CO$_2$. However, our carbon model suggests that such states would not have been stable. Indeed, we found that warm climates can be maintained only if the seafloor weathering is assumed independent of temperature. Such an assumption is inconsistent with recent measurements indicating a higher weathering during the late Mesozoic, when the deep ocean was around 10$^\circ$C warmer \citep{coogan13}. Stable very warm climates with a N$_2$-CO$_2$-CH$_4$-H$_2$O early atmosphere are therefore excluded.

Our results favor instead temperate climates with global mean surface temperatures around 8.5-30.5$^\circ$C and with pCO$_2$ around 0.1-0.36 bar. With the increasing solar luminosity over the Archean, the carbon cycle would have led to a reduction of pCO$_2$. In parallel, with the emergence of land and the slowing seafloor spreading rate during the Archean, the carbon cycle would have progressively become controlled by continental weathering while initially controlled by seafloor weathering. 
However, a better understanding of the data from O isotopes in ancient cherts together with better constraints on pCO$_2$ are required to validate our conclusions about the long-debated climate of early Earth.

During the Late Heavy Bombardment, the weathering of ejecta would have dramatically decreased the partial pressure of CO$_2$ leading to cold climates but not necessary to a snowball Earth during all that period. 
We predict that large impactors (i.e. with diameters higher than around 50 km) could have trigger extreme glaciations during the Hadean and the Archean.

Finally, our modeling work suggests that the CO$_2$-dependent carbon cycle could have maintained temperate climates on the early Earth, without requiring any additional greenhouse gas or warming process. This work also highlights the importance of the seafloor weathering, which can dominate the continental weathering. Under such circumstances, the climate becomes relatively insensitive to the presence of land. 
The strong temperature dependence expected for the seafloor weathering also has implications for the habitability of exoplanets, allowing in particular a stabilizing climate-weathering feedback on waterworlds. Waterworlds might therefore not be much more subject to water loss by moist greenhouse than partially ocean-covered Earth-like planets, as suggested by \cite{abbot12}.
\\
\\
\paragraph{Acknowledgments:}
B.C. acknowledges support from Paris Sciences et Lettres and from an appointment to the NASA Postdoctoral Program, administered by USRA.
This work was performed as part of the NASA Astrobiology Institute's Virtual Planetary Laboratory, supported by NASA under Cooperative Agreement No. NNA13AA93A.  DC also acknowledges support from NASA Exobiology Program grant NNX15AL23G.
We thank J. Kasting and an anonymous reviewer for helpful comments.

\appendix

\section{Carbon cycle modeling}

\begin{figure}[h] 
\begin{center} 
	\includegraphics[width=15cm]{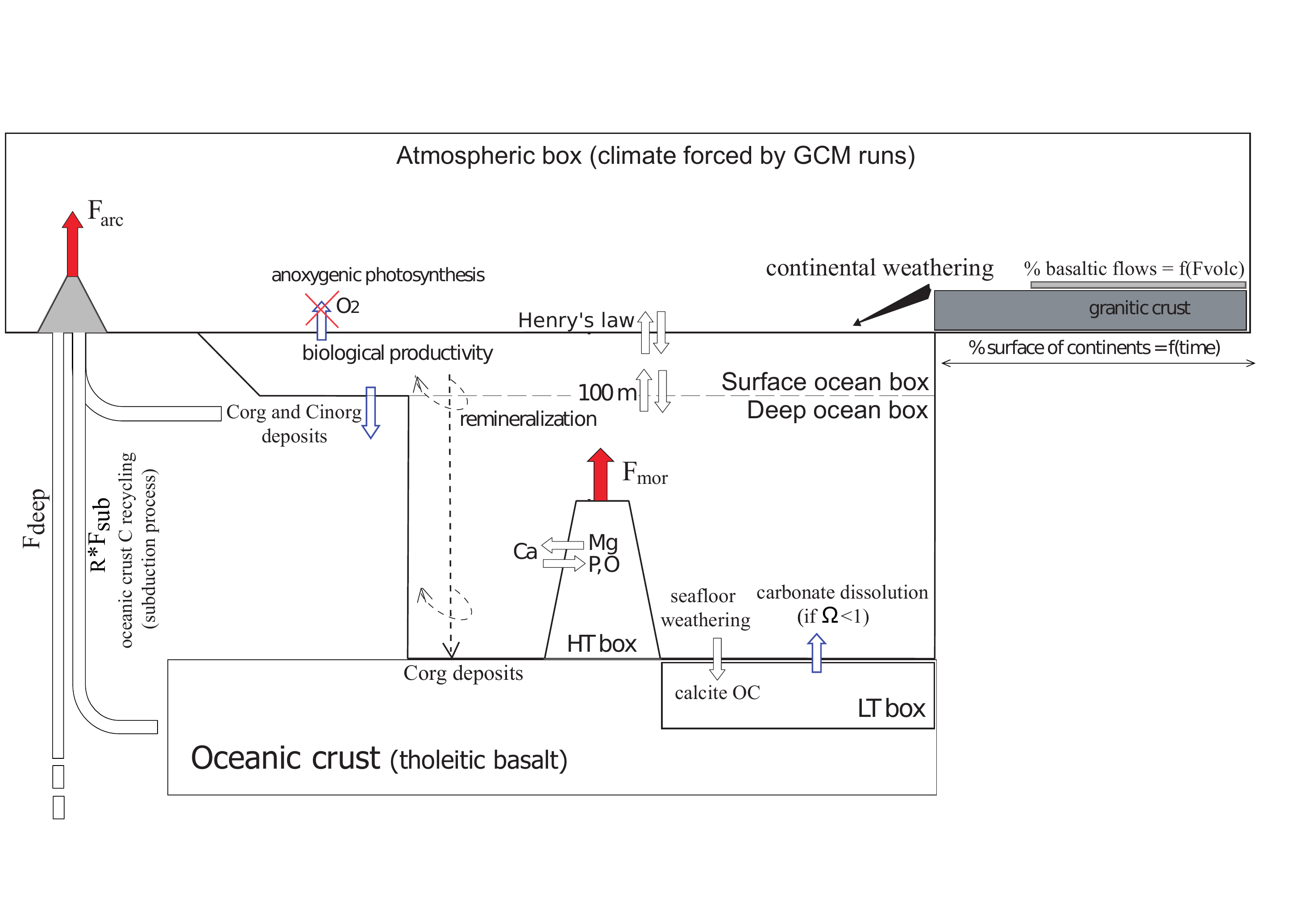}
\end{center} 
\caption{Representation of the carbon model used for the Hadean Earth.} 
\label{figure_S1}
\end{figure} 

The carbon cycle model has been previously used for the Proterozoic and the Phanerozoic \citep{godderis06, lehir08a, lehir08b}. Figure \ref{figure_S1} shows the different processes included in the model. The parametrizations for the different CO$_2$ flux are briefly described below.
\\
\\
\textbf{\textit{{CO$_2$ outgassing by oceanic ridges and arc volcanoes}}}
\\
We considered that the CO$_2$ flux from mid oceanic ridges $F_{mor}$ is proportional to the production of oceanic crust:

\begin{eqnarray}
\it F_{ \rm mor}=F_{\rm mor_0} \times \frac{SR}{SR_0} \times \frac{H}{H_0} 
\end{eqnarray}
\\
where $F_{mor_0}$,= 1.8$\times$10$^{12}$ mol/yr, $SR$ is the spreading rate and $H$ is the oceanic crust thickness ($H_0$ = 6 km).

We assumed that the flux from arc volcanoes (F$_{\rm arc}$) is the total degassing flux due to carbonate recycling at subduction zones and mantle outgassing (with the exception of mid-oceanic ridge). Degassing over geologic times being assumed proportional to the heat flux from the interior of the Earth, the mantle outgassing rate is adjusted, as appropriate, using a forcing factor, rdeg=2.5 at 3.8 Ga \citep{Godderis00}):

\begin{eqnarray}
\it F_{\rm arc} = F_{\rm sub} \times R + F_{deep}
\end{eqnarray}
\\
where $R$ is the recycling rate and $F_{deep}=C_{\rm deep}^0 \times rdeg$ with $C_{deep}^0$ = 3.3 $\times$ 10$^{12}$ mol/yr.
For the heat-pipe case, we considered that $SR$ = 0, $F_{\rm sub}$ = 0 and $rdeg$ =10. Indeed, according to \cite{moore13}, the heat flux had to be around 10 times higher than today to be in the heat-pipe regime.
\\
\\
\textbf{\textit{{Continental weathering}}}
\\
Weathering depends on rocks available for reactions (basalt vs. granite) and fraction of continents deduced from crustal growth models. 
The weathering rate of continental granites and basalts is given by:

\begin{eqnarray}
\it F_{\rm con} =f_{\rm Ca}^{bas} + f_{\rm Ca}^{gra}+f_{\rm Mg}^{bas} + f_{\rm Mg}^{gra}
\end{eqnarray}
\\
with:
\\
\begin{eqnarray}
\it f_{\rm Ca}^{bas} = 20.94\times10^{12} \times e^{fbas} \times runoff \times  \frac{srf_{\rm bas}}{srf_{\rm cont}} \times \frac{srf_{\rm cont}}{srf_{\rm cont0}}  \\
f_{\rm  Ca} ^{gra} = 3.49 \times 10^{12} \times e^{fgra} \times runoff \times  \left(1-\frac{srf_{\rm bas}}{srf_{\rm cont}}\right) \times \frac{srf_{\rm cont}}{srf_{\rm cont0}}  \\
f_{\rm Mg}^{bas} = 19.86 \times 10^{12} \times e^{fbas} \times runoff \times  \frac{srf_{\rm bas}}{srf_{\rm cont}} \times \frac{srf_{\rm cont}}{srf_{\rm cont0}}  \\
f_{\rm Mg}^{gra} = 3.31 \times 10^{12} \times e^{fgra} \times runoff \times  \left(1-\frac{srf_{\rm bas}}{srf_{\rm cont}}\right) \times \frac{srf_{\rm cont}}{srf_{\rm cont0}}
\end{eqnarray}
\\
with $fgra=\frac{-48200}{8.314} \left(\frac{1}{T}-\frac{1}{T_0} \right)$, $fbas=\frac{-42300}{8.314} \left(\frac{1}{T}-\frac{1}{T_0} \right)$ and where $\frac{srf_{\rm cont}}{srf_{\rm cont0}}$ is the fraction of emerged continental area.
\\
We used $pCO_2^0$ = 285 ppm. 
These values give a present-day flux of 3.49$\times$10$^{12}$ mol/yr for Ca$^{2+}$ and 3.31$\times$10$^{12}$ mol/yr for Mg$^{2+}$.

To represent the presence of very active volcanoes and potential extent of basaltic flows over small continents, rocks available for weathering is assumed proportional to volcanism ( $F_{\rm arc}$).
Based on the extent of basalt outcrops of the Cenozoic era, the equilibrium between basaltic flows / volcanism is prescribed as follow: 
\begin{eqnarray}
\it srf_{\rm bas} = \frac{F_{\rm arc}}{F_{\rm arc0}} \times \frac{srf_{\rm cont}}{srf_{\rm cont0}} \times srf_{\rm bas0}
\end{eqnarray}
with $F_{\rm arc0} = 6.8 \times 10^{12}$ mol/yr \citep{gaillardet99}, $srf_{cont0}  = 146 \times 10^6$ km$^2$ and $srf_{bas0}  = 7.3 \times 10^6$ km$^2$ \citep{dessert03}.

Using this assumption, our model predicts that early Archean continents were mostly mafic, composed ~ 65-70$\%$ of basalts and 35-30$\%$ of granite, consistent with analysis of Sb/Sr, Ni/Co and Cr/Zn ratios by \cite{dhuime15, tang16}.

The runoff and the surface temperature depend on the latitude and are properly computed by the GCM. We assumed that lands are homogenously distributed over all the Earth. Then, we computed the global continental weathering flux by summing the contribution of each latitudinal band from the GCM.
\\
\\
\textbf{\textit{{Seafloor weathering}}}
\\
Contrary to \cite{caldeira95}, we assume that the seafloor weathering does not occur in solutions at steady state with carbonate minerals, but when oceanic water enters in contact with fresh basaltic rock, it evolves along its downward path towards saturation with calcite, which is the ultimate state of the reaction. Our percolating reservoir stands for water during the early stages of seafloor weathering, when all cations are produced. The seafloor weathering rate is given by \citep{godderis06, lehir08a, lehir08b}:

\begin{eqnarray}
\it f_{sfw}=\frac{SR}{SR_0} \times \sum\limits_{i} d_i exp\left(\frac{-E_i}{R T_p} \right) C_i^n
\end{eqnarray}
\\
where $SR$ is the spreading rate. The sum extends up to all the dissolving species (H$^+$, OH$^–$ and H$_2$O), while the index $i$ stands for the dissolving species. $E_i$ and $d_i$ are, respectively, the activation energy and dissolution constant for each mineral depending on the species $i$ promoted dissolution. $T_p$ is the temperature of the percolating waters into the oceanic crust at which dissolution occurs. We divided the oceanic crust in 6 layers from the seafloor to a depth of 500 m in the crust. We assumed that the percolating water has the temperature of surrounding crust fixed with the present-day temperature gradient (around 116 K/km). For the first layer, the deep ocean temperature is assumed to be equal to the sea surface temperature at 60$^{\circ}$ N/S (see idealized simulations by \cite{manabe85, enderton09}). We scaled the seafloor weathering to its present-day value.
$C_i$ are the concentration in the species $i$, while $n_i$ is the order of the dissolution reaction.
We considered that the seafloor weathering occurs only in the first 500 m of the oceanic crust. A change of oceanic crust thickness thus does not impact the seafloor weathering, which remains proportional to the spreading rate.

Experimental studies and measurements of carbonate mineral in the oceanic crust reveal that the seafloor weathering strongly depend on the bottom ocean temperature and was around 5 times higher during the late Mesozoic \citep{brady97, coogan13}, when the deep ocean temperature was around $\sim$10 K higher. It is therefore expected to play a significant role on the early Earth with a lower land cover and a faster spreading rate.
Our model predict a 2-3 times higher seafloor weathering during the late Mesozoic for 4 times the present-day pCO$_2$. 
However, the volcanic outgassing may have been higher and continental weathering may have been less efficient at that time, what would have increased the seafloor weathering closer to the measured rates. Because of these uncertainties, we consider that our values remain acceptable but we may underestimate the temperature sensitivity of the seafloor weathering.
\\
\\
\textbf{\textit{{Precipitation of carbonates}}}
\\
In the absence of pelagic producers, carbonates are assumed to be precipitated on continental margins (at a depth of less than 100 m) through: 

\begin{eqnarray}
\it f_{cd}=k_{cd}(\Omega-1)^{1.7}
\end{eqnarray}
\\
where $k_{cd}$ = 0.172 mol/yr/m$^2$ and $\Omega$ is the saturation ratio of calcite: $\Omega$ = $\frac{[Ca^{2+}][CO_3^{2-}]}{K_s}$ where $K_s$ is the solubility constant depending on the temperature and the salinity (expression given in \cite{sarmiento06}).
\\
\\
\textbf{\textit{{Ocean-atmosphere interface}}}
\\
The CO$_2$ exchange flux between the atmosphere and the oceanis given by: 

\begin{eqnarray}
\it f_{atm-ocean}=\alpha (pCO_2^{atm}-pCO_2^{ocean})
\end{eqnarray}
where $\alpha$ = 16 mol/m$^2$/PAL and $pCO_2^{ocean}$ is the CO$_2$ partial pressure computed with Henry's law for the concentration of dissolved CO$_2$.
\\
\\
\textbf{\textit{{Seafloor spreading rate}}}
\\
We computed the seafloor spreading rate $SR$ considering it evolves with the potential temperature of the mantle $T_p$ as Rayleigh-B\'enard convection  \citep{flament08}:
\begin{eqnarray}
\it SR = SR_* \times \left( \frac{T_p-T_0}{T_p^*-T_0} \frac{\eta(T_p^*)}{ \eta(T_p)}\right)^{2/3}
\end{eqnarray}
where $T_0$ is the deep ocean temperature (we assumed $T_0$ = 0$^\circ$C, but the value has a very small impact on $SR$), $\eta(T_p)=\eta_0 \left(\frac{E}{R T_p} \right)$ is the viscosity of the mantle with $E$ = 380 kJ and $R$ = 8.31 J/K/mol. Assuming $T_p^*$=1330$^{\circ}$C for the present-day Earth and $T_p$=1550$^\circ$C for the early Earth then gives $u/u_*$ = 11. If the oceanic crust thickness is increased by a factor 3, then the spreading rate becomes 3 times lower while the crust production rate and the heat flux remain constant. A spreading rate 11 times higher than today gives a lifetime of the oceanic crust of around 15 Ma.

\section{Effects of impacts during the Late Heavy Bombardment}

\textbf{\textit{{Ejecta weathering}}}
\\
We computed the effect of ejecta weathering following \cite{sleep01}. Impacts produce a global layer of particles with diameters typically from 1 mm to 100 m. Ejecta constitute a source of cations that precipitate with carbonate ions.
The mass of ejecta was determined using the scaling laws from \cite{collins05}. We took into account the deceleration caused by the atmosphere and the ocean but we neglected the breakup in the atmosphere and the mass loss by sublimation. These assumptions are valid for large impactors. We also included the escaping mass of impactor after the impact according to the hydrodynamic simulations from \cite{shuvalov09}.

We assumed that impactor have a velocity $v_0$ of 21 km/s for asteroids and 30 km/s for comets, a diameter $L$, an impact angle $\theta$ of 45$^\circ$ and a density $\rho_i$ = 3000 kg /m$^3$ for asteroids and $\rho_i$ = 1000 kg /m$^3$ for comets. The Earth is supposed covered by a 2700 m deep global ocean. We chose a density $\rho_c$ = 2700 kg/m$^3$ for the crust.
\\
The mass of ejecta is computed according to \cite{collins05}:

\begin{eqnarray}
\it m_{ej}=\frac{\pi D_{tc}^3 \rho_{c}}{16\sqrt 2}
\end{eqnarray}
\\
where the transient crater $D_{tc}$ (in m) is given by \citep{collins05}:

\begin{eqnarray}
\it D_{tc}=1.161\left( \frac{\rho_i}{\rho_c}\right)^{1/3} L^{0.78} v_i^{0.44} g_E^{-0.22} sin^{1/3}\theta
\end{eqnarray}
\\
where $g_E$ is Earth gravity and $v_i$ is the impactor velocity at the seafloor.
\\
Taking into account the deceleration by the atmosphere and the ocean, the impactor velocity at the seafloor is \citep{collins05}:

\begin{eqnarray}
\it v_{i}=v_{0} exp\left( -\frac{3 (C_{Datm} \rho_{air}  H_{atm} + C_{Docean} \rho_{water} H_{ocean})}{4\rho_i L sin \theta}\right)
\end{eqnarray}
\\
where $C_D$ is the drag coefficient, taken equal to 2 for the atmosphere and 0.877 for the ocean, $H_{atm}$ is the atmospheric scale height, and $H_{ocean}$ is the ocean depth.
For a 10 km diameter impactor, the atmosphere reduces the velocity by just 0.1$\%$ but the ocean reduces it by around 8$\%$.

\cite{collins05} suggest that the mean diameter of ejecta decreases with distance from the crater center $r$ as $d \propto r^{-\alpha}$, where $\alpha$=2.65 and that the thickness of the ejecta layer decreases as $h \propto r^{-3}$. By integrating these relations, the mass distribution of ejecta is given by $\it N(\geq m) \propto m^{-\gamma}$, where $\gamma$=0.87. \cite{zahnle02} use the same law with $\gamma$=0.9-0.95.
We assumed that the diameter of each particle decreases of 1 mm in 167 kyrs \citep{crovisier87} and that cations are present at 0.005 mol/g.

With the impact statistics that we use, the impact flux is around $6.7\times 10^{11}$ kg/yr and the ejecta production rate is around $3.2\times 10^{12}$ kg/yr, corresponding to a cation flux of $1.6 \times 10^{13}$ mol/yr (compared to 3 $\times 10^{14}$ mol/yr in \cite{sleep01}).
In our carbon cycle model, these cations precipitate if the saturation ratio of calcite is higher than 1 (see Appendix A).
\\
\\
\textbf{\textit{{CO$_2$ outgassing by impacts}}}
\\
We assumed that the early oceanic crust is covered by a 40 m thick carbonate layer ($\rho_{carb}$ = 1500 kg/m$^3$). We therefore assumed the equivalent of 4 bars of CO$_2$ ($4.8\times 10^{20}$ moles of CO$_2$) are in this layer and can be released during impacts. This amount is lower than the total quantity of CO$_2$ currently stored in the oceanic crust (around $1.2\times 10^{21}$ moles of CO$_2$ \citep{sleep01}).
The shock wave produced by an impact can lead to decarbonation of carbonates. For calcite, the reaction is CaCO$_3$ $\Rightarrow$ CaO + CO$_2$. 
This process starts when the pressure is higher than 20 GPa and is total when the pressure is higher than 60 GPa \citep{agrinier01}. However, the amount of outgassed CO$_2$ is reduced by recombination through the back reaction. The percentage of CO$_2$ back reacted is between 37 and 66 $\%$ \citep{agrinier01}. For simplicity, we assumed that CO$_2$ outgassing occurs only at pressures higher than 60 GPa with an efficiency of 50 $\%$ due to the reverse reaction.
For each impact, we computed the area where the shock pressure is higher than 60 GPa, assuming that it evolves with the distance $r$ to the impact center as $\propto r^{-1.5}$.

We also took into account CO$_2$ outgassed by melting lithosphere, assuming that its volume $V_m$ is \citep{collins05}:

\begin{eqnarray}
\it V_{m}=8.9 \times10^{-12} E sin\theta
\end{eqnarray}
\\
where $E$ is the impactor kinetic energy and $\theta$=45$^\circ$ is the impact angle. We assumed that the lithosphere below the carbonate layer contains 100 ppmv of CO$_2$ and outgasses all CO$_2$ when melted.



\newpage
\bibliographystyle{agu}

\begin{thebibliography}{73}
\providecommand{\natexlab}[1]{#1}
\expandafter\ifx\csname urlstyle\endcsname\relax
  \providecommand{\doi}[1]{doi:\discretionary{}{}{}#1}\else
  \providecommand{\doi}{doi:\discretionary{}{}{}\begingroup
  \urlstyle{rm}\Url}\fi

\bibitem[{\textit{{Abbot} et~al.}(2012)\textit{{Abbot}, {Cowan}, and
  {Ciesla}}}]{abbot12}
{Abbot}, D.~S., N.~B. {Cowan}, and F.~J. {Ciesla}, {Indication of Insensitivity
  of Planetary Weathering Behavior and Habitable Zone to Surface Land Fraction}, \textit{Astrophysical Journal}, \textit{756}, 178,
  \doi{10.1088/0004-637X/756/2/178}, 2012.

\bibitem[{\textit{{Abramov} and {Mojzsis}}(2009)}]{abramov09}
{Abramov}, O., and S.~J. {Mojzsis}, {Microbial habitability of the Hadean Earth
  during the late heavy bombardment}, \textit{Nature}, \textit{459}, 419--422,
  \doi{10.1038/nature08015}, 2009.

\bibitem[{\textit{{Agrinier} et~al.}(2001)\textit{{Agrinier}, {Deutsch},
  {Sch{\"a}rer}, and {Martinez}}}]{agrinier01}
{Agrinier}, P., A.~{Deutsch}, U.~{Sch{\"a}rer}, and I.~{Martinez}, {Fast
  back-reactions of shock-released CO $_{2}$ from carbonates: an experimental
  approach}, \textit{Geochimica et Cosmochimica Acta}, \textit{65}, 2615--2632,
  \doi{10.1016/S0016-7037(01)00617-2}, 2001.

\bibitem[{\textit{{Anbar} et~al.}(2001)\textit{{Anbar}, {Zahnle}, {Arnold}, and
  {Mojzsis}}}]{anbar01}
{Anbar}, A.~D., K.~J. {Zahnle}, G.~L. {Arnold}, and S.~J. {Mojzsis},
  Extraterrestrial iridium, sediment accumulation and the habitability of the
  early earth's surface, \textit{Journal of Geophysical Research}, \textit{106}, 3219--3236,
  \doi{10.1029/2000JE001272}, 2001.

\bibitem[{\textit{{Bell} et~al.}(2015)\textit{{Bell}, {Boehnke}, {Harrison},
  and {Mao}}}]{bell15}
{Bell}, E., P.~{Boehnke}, T.~M. {Harrison}, and W.~L. {Mao}, Potentially
  biogenic carbon preserved in a 4.1 billion-year-old zircon,
  \textit{Proceedings of the National Academy of Sciences}, \textit{112}(47),
  14,518--14,521, \doi{10.1073/pnas.1517557112}, 2015.

\bibitem[{\textit{{Bickle}}(1986)}]{bickle86}
{Bickle}, M.~J., {Implications of melting for stabilisation of the lithosphere
  and heat loss in the Archaean}, \textit{Earth and Planetary Science Letters},
  \textit{80}, 314--324, \doi{10.1016/10.1016/0012-821X(86)90113-5}, 1986.

\bibitem[{\textit{{Blake} et~al.}(2010)\textit{{Blake}, {Chang}, and
  {Lepland}}}]{blake10}
{Blake}, R.~E., S.~J. {Chang}, and A.~{Lepland}, {Phosphate oxygen isotopic
  evidence for a temperate and biologically active Archaean ocean},
  \textit{Nature}, \textit{464}, 1029--1032, \doi{10.1038/nature08952}, 2010.

\bibitem[{\textit{{Boehnke} and {Harrison}}(2016)}]{Boehnke16}
{Boehnke}, P., and T.~M. {Harrison}, {Illusory Late Heavy Bombardments},
  \textit{Proceedings of the National Academy of Science}, \textit{113},
  10,802--10,806, \doi{10.1073/pnas.1611535113}, 2016.

\bibitem[{\textit{{Bottke} and {Norman}}(2017)}]{bottke17}
{Bottke}, W.~F., and M.~D. {Norman}, {The Late Heavy Bombardment},
  \textit{Annual Reviews of Earth and Planetary Sciences}, \textit{45},
  \doi{10.1146/annurev-earth-063016-020131}, 2017.

\bibitem[{\textit{{Bottke} et~al.}(2012)\textit{{Bottke}, {Vokrouhlick{\'y}},
  {Minton}, {Nesvorn{\'y}}, {Morbidelli}, {Brasser}, {Simonson}, and
  {Levison}}}]{bottke12}
{Bottke}, W.~F., D.~{Vokrouhlick{\'y}}, D.~{Minton}, D.~{Nesvorn{\'y}},
  A.~{Morbidelli}, R.~{Brasser}, B.~{Simonson}, and H.~F. {Levison}, {An
  Archaean heavy bombardment from a destabilized extension of the asteroid
  belt}, \textit{Nature}, \textit{485}, 78--81, \doi{10.1038/nature10967},
  2012.

\bibitem[{\textit{{Bousseau} et~al.}(2008)\textit{{Bousseau}, {Blanquart},
  {Necsulea}, {Lartillot}, and {Manolo}}}]{boussau08}
{Bousseau}, B., S.~{Blanquart}, A.~{Necsulea}, N.~{Lartillot}, and G.~{Manolo},
  {Parallel adaptations to high temperatures in the Archaean eon},
  \textit{Nature}, \textit{456}, 942--945, \doi{10.1038/nature07393}, 2008.

\bibitem[{\textit{{Brady} and {G{\'{\i}}slason}}(1997)}]{brady97}
{Brady}, P.~V., and S.~R. {G{\'{\i}}slason}, {Seafloor weathering controls on
  atmospheric CO $_{2}$ and global climate}, \textit{Geochimica et Cosmochimica
  Acta}, \textit{61}, 965--973, \doi{10.1016/S0016-7037(96)00385-7}, 1997.

\bibitem[{\textit{{Buick} et~al.}(1981)\textit{{Buick}, {Dunlop}, and
  {Groves}}}]{buick81}
{Buick}, R., J.~{Dunlop}, and D.~{Groves}, Stromatolite recognition in ancient
  rocks: an appraisal of irregularly laminated structures in an early archaean
  chert-barite unit from north pole, western australia, \textit{Alcheringa: An
  Australasian Journal of Palaeontology}, \textit{5}(3), 161--181,
  \doi{10.1080/03115518108566999}, 1981.

\bibitem[{\textit{{Caldeira}}(1995)}]{caldeira95}
{Caldeira}, K., {Long-term control of atmospheric carbon dioxide:
  Low-temperature seafloor alteration or terrestrial silicate-rock
  weathering?}, \textit{American Jounral of Science}, \textit{295}, 1077--1114,
  \doi{10.2475/ajs.295.9.1077}, 1995.

\bibitem[{\textit{{Charnay} et~al.}(2013)\textit{{Charnay}, {Forget},
  {Wordsworth}, {Leconte}, {Millour}, {Codron}, and {Spiga}}}]{charnay13}
{Charnay}, B., F.~{Forget}, R.~{Wordsworth}, J.~{Leconte}, E.~{Millour},
  F.~{Codron}, and A.~{Spiga}, {Exploring the faint young Sun problem and the
  possible climates of the Archean Earth with a 3-D GCM}, \textit{Journal of Geophysical Research},
  \textit{118}, 10,414, \doi{10.1002/jgrd.50808}, 2013.

\bibitem[{\textit{{Codron}}(2012)}]{codron12}
{Codron}, F., {Ekman heat transport for slab oceans}, \textit{Climate
  Dynamics}, \textit{38}, 379--389, \doi{10.1007/s00382-011-1031-3}, 2012.

\bibitem[{\textit{{Collins} et~al.}(2005)\textit{{Collins}, {Melosh}, and
  {Marcus}}}]{collins05}
{Collins}, G.~S., H.~J. {Melosh}, and R.~A. {Marcus}, {Earth Impact Effects
  Program: A Web-based computer program for calculating the regional
  environmental consequences of a meteoroid impact on Earth},
  \textit{Meteoritics and Planetary Science}, \textit{40}, 817,
  \doi{10.1111/j.1945-5100.2005.tb00157.x}, 2005.

\bibitem[{\textit{{Coogan} and {Gillis}}(2013)}]{coogan13}
{Coogan}, L.~A., and K.~M. {Gillis}, {Evidence that low-temperature oceanic
  hydrothermal systems play an important role in the silicate-carbonate
  weathering cycle and long-term climate regulation}, \textit{Geochemistry,
  Geophysics, Geosystems}, \textit{14}, 1771--1786, \doi{10.1002/ggge.20113},
  2013.

\bibitem[{\textit{{Crovisier} et~al.}(1987)\textit{{Crovisier}, {Honnorez}, and
  {Eberhart}}}]{crovisier87}
{Crovisier}, J.~L., J.~{Honnorez}, and J.~P. {Eberhart}, {Dissolution of
  basaltic glass in seawater: Mechanism and rate}, \textit{Geochimica et
  Cosmochimica Acta}, \textit{51}, 2977--2990,
  \doi{10.1016/0016-7037(87)90371-1}, 1987.

\bibitem[{\textit{{de Niem} et~al.}(2012)\textit{{de Niem}, {K{\"u}hrt},
  {Morbidelli}, and {Motschmann}}}]{deniem12}
{de Niem}, D., E.~{K{\"u}hrt}, A.~{Morbidelli}, and U.~{Motschmann},
  {Atmospheric erosion and replenishment induced by impacts upon the Earth and
  Mars during a heavy bombardment}, \textit{Icarus}, \textit{221}, 495--507,
  \doi{10.1016/j.icarus.2012.07.032}, 2012.

\bibitem[{\textit{de~Wit and Furnes}(2016)}]{dewit16}
de~Wit, M.~J., and H.~Furnes, 3.5-ga hydrothermal fields and diamictites in the
  barberton greenstone belt{\textemdash}paleoarchean crust in cold
  environments, \textit{Science Advances}, \textit{2}(2),
  \doi{10.1126/sciadv.1500368}, 2016.

\bibitem[{\textit{{Dessert} et~al.}(2003)\textit{{Dessert}, {Gaillardet}, L.,
  and C.}}]{dessert03}
{Dessert}, B., C.and~{Dupr{\'e}}, J.~{Gaillardet}, F.~L., and A.~C., {Basalt
  weathering laws and the impact of basalt weathering on the global carbon
  cycle}, \textit{Chem. Geol}, \textit{202}, 257--273, 2003.

\bibitem[{\textit{{Dhuime} et~al.}(2012)\textit{{Dhuime}, {Hawkesworth},
  {Cawood}, and {Storey}}}]{dhuime12}
{Dhuime}, B., C.~J. {Hawkesworth}, P.~A. {Cawood}, and C.~D. {Storey}, {A
  Change in the Geodynamics of Continental Growth 3 Billion Years Ago},
  \textit{Science}, \textit{335}, 1334--, \doi{10.1126/science.1216066}, 2012.

\bibitem[{\textit{{Dhuime} et~al.}(2015)\textit{{Dhuime}, {Wuestefeld}, and
  {Hawkesworth}}}]{dhuime15}
{Dhuime}, B., A.~{Wuestefeld}, and C.~J. {Hawkesworth}, {Emergence of modern
  continental crust about 3 billion years ago}, \textit{Nature Geoscience},
  \textit{8}, 552--555, \doi{10.1038/ngeo2466}, 2015.

\bibitem[{\textit{{Driese} et~al.}(2011)\textit{{Driese}, {Jirsa}, {Ren},
  {Brantley}, {Sheldon}, {Parker}, and {Schmitz}}}]{driese11}
{Driese}, S., M.~{Jirsa}, M.~{Ren}, S.~{Brantley}, N.~{Sheldon}, D.~{Parker},
  and M.~{Schmitz}, Neoarchean paleoweathering of tonalite and metabasalt:
  Implications for reconstructions of 2.69 ga early terrestrial ecosystems and
  paleoatmospheric chemistry, \textit{Precambrian Research}, \textit{189},
  1--17, \doi{10.1016/j.precamres.2011.04.003}, 2011.

\bibitem[{\textit{{Enderton} and {Marshall}}(2009)}]{enderton09}
{Enderton}, D., and J.~{Marshall}, {Explorations of Atmosphere-Ocean-Ice
  Climates on an Aquaplanet and Their Meridional Energy Transports},
  \textit{Journal of Atmospheric Sciences}, \textit{66}, 1593--1611,
  \doi{10.1175/2008JAS2680.1}, 2009.

\bibitem[{\textit{{Flament} et~al.}(2008)\textit{{Flament}, {Coltice}, and
  {Rey}}}]{flament08}
{Flament}, N., N.~{Coltice}, and P.~F. {Rey}, {A case for late-Archaean
  continental emergence from thermal evolution models and hypsometry},
  \textit{Earth and Planetary Science Letters}, \textit{275}, 326--336,
  \doi{10.1016/j.epsl.2008.08.029}, 2008.

\bibitem[{\textit{Forget et~al.}(2013)\textit{Forget, Wordsworth, Millour,
  Madeleine, Kerber, Leconte, Marcq, and Haberle}}]{forget13}
Forget, F., R.~Wordsworth, E.~Millour, J.~B. Madeleine, L.~Kerber, J.~Leconte,
  E.~Marcq, and R.~M. Haberle, {3D modelling of the early martian climate under
  a denser CO2 atmosphere: Temperatures and CO2 ice clouds}, \textit{Icarus},
  \textit{222}(1), 81--99, \doi{10.1016/j.icarus.2012.10.019}, 2013.

\bibitem[{\textit{{Fralick} and {Carter}}(2011)}]{fralick11}
{Fralick}, P., and J.~{Carter}, {Neoarchean deep marine paleotemperature:
  Evidence from turbidite successions}, \textit{Precambrian Research},
  \textit{191}, 78--84, \doi{10.1016/j.precamres.2011.09.004}, 2011.

\bibitem[{\textit{{Gaillardet} et~al.}(1999)\textit{{Gaillardet}, {Dupr\'e},
  P., and J.}}]{gaillardet99}
{Gaillardet}, J., B.~{Dupr{\'e}}, L.~P., and A.~C. J., {Global silicate
  weathering of silicates estimated from large river geochemistry},
  \textit{Chem. Geol., Special issue Carbon Cycle}, \textit{7159}, 3--30, 1999.

\bibitem[{\textit{{Gaucher} et~al.}(2008)\textit{{Gaucher}, {Govindarajan}, and
  {Ganesh}}}]{gaucher08}
{Gaucher}, E.~A., S.~{Govindarajan}, and O.~K. {Ganesh}, {Palaeotemperature
  trend for Precambrian life inferred from resurrected proteins},
  \textit{Nature}, \textit{452}, 704--707, \doi{10.1038/nature06510}, 2008.

\bibitem[{\textit{{Godd{\'e}ris} and {Veizer}}(2000)}]{Godderis00}
{Godd{\'e}ris}, Y., and J.~{Veizer}, {Tectonic control of chemical and isotopic
  composition of ancient oceans; the impact of continental growth},
  \textit{American Journal of Sciences}, \textit{300}, 434--461, \doi{doi:
  10.2475/ajs.300.5.434}, 2000.

\bibitem[{\textit{{Godd{\'e}ris} et~al.}(2006)\textit{{Godd{\'e}ris}, {Fran{\c
  c}ois}, {Probst}, {Schott}, {Moncoulon}, {Labat}, and
  {Viville}}}]{godderis06}
{Godd{\'e}ris}, Y., L.~M. {Fran{\c c}ois}, A.~{Probst}, J.~{Schott},
  D.~{Moncoulon}, D.~{Labat}, and D.~{Viville}, {Modelling weathering processes
  at the catchment scale: The WITCH numerical model}, \textit{Geochimica et
  Cosmochimica Acta}, \textit{70}, 1128--1147, \doi{10.1016/j.gca.2005.11.018},
  2006.

\bibitem[{\textit{{Gomes} et~al.}(2005)\textit{{Gomes}, {Levison}, {Tsiganis},
  and {Morbidelli}}}]{gomes05}
{Gomes}, R., H.~F. {Levison}, K.~{Tsiganis}, and A.~{Morbidelli}, {Origin of
  the cataclysmic Late Heavy Bombardment period of the terrestrial planets},
  \textit{Nature}, \textit{435}, 466--469, \doi{10.1038/nature03676}, 2005.

\bibitem[{\textit{{Haqq-Misra} et~al.}(2016)\textit{{Haqq-Misra}, {Kopparapu},
  {Batalha}, {Harman}, and {Kasting}}}]{haqq-misra16}
{Haqq-Misra}, J., R.~K. {Kopparapu}, N.~E. {Batalha}, C.~E. {Harman}, and J.~F.
  {Kasting}, {Limit Cycles Can Reduce the Width of the Habitable Zone},
  \textit{Astrophys. J.}, \textit{827}, 120, \doi{10.3847/0004-637X/827/2/120},
  2016.

\bibitem[{\textit{{Hoareau} et~al.}(2015)\textit{{Hoareau}, {Bomou}, {van
  Hinsbergen}, {Carry}, {Marquer}, {Donnadieu}, {Le Hir}, {Vrielynck}, and
  {Walter-Simonnet}}}]{hoareau15}
{Hoareau}, G., B.~{Bomou}, D.~J.~J. {van Hinsbergen}, N.~{Carry}, D.~{Marquer},
  Y.~{Donnadieu}, G.~{Le Hir}, B.~{Vrielynck}, and A.-V. {Walter-Simonnet},
  {Did high Neo-Tethys subduction rates contribute to early Cenozoic warming?},
  \textit{Climate of the Past}, \textit{11}, 1751--1767,
  \doi{10.5194/cp-11-1751-2015}, 2015.

\bibitem[{\textit{{Hren} et~al.}(2009)\textit{{Hren}, {Tice}, and
  {Chamberlain}}}]{hren09}
{Hren}, M.~T., M.~M. {Tice}, and C.~P. {Chamberlain}, {Oxygen and hydrogen
  isotope evidence for a temperate climate 3.42 billion years ago},
  \textit{Nature}, \textit{462}, 205√ê208, \doi{10.1038/nature08518}, 2009.

\bibitem[{\textit{{Johnson} and {Melosh}}(2012)}]{johnson12}
{Johnson}, B.~C., and H.~J. {Melosh}, {Impact spherules as a record of an
  ancient heavy bombardment of Earth}, \textit{Nature}, \textit{485}, 75--77,
  \doi{10.1038/nature10982}, 2012.

\bibitem[{\textit{{Kanzaki} and {Murakami}}(2015)}]{kanzaki15}
{Kanzaki}, Y., and T.~{Murakami}, {Estimates of atmospheric CO$_{2}$ in the
  Neoarchean-Paleoproterozoic from paleosols}, \textit{Geochimica et
  Cosmochimica Acta}, \textit{159}, 190--219, \doi{10.1016/j.gca.2015.03.011},
  2015.

\bibitem[{\textit{{Kasting} and {Howard}}(2006)}]{kasting06a}
{Kasting}, J.~F., and T.~M. {Howard}, {Atmospheric composition and climate on
  the early Earth }, \textit{Phil. Trans. R. Soc.}, \textit{361}, 1733--1742,
  \doi{10.1098/rstb.2006.1902}, 2006.

\bibitem[{\textit{{Kharecha} et~al.}(2005)\textit{{Kharecha}, {Kasting}, and
  {Siefert}}}]{kharecha05}
{Kharecha}, P., J.~{Kasting}, and J.~{Siefert}, A coupled
  atmosphere-ecosystem model of the early archean earth,
  \textit{Geobiology}, \textit{3}(2), 53--76,
  \doi{10.1111/j.1472-4669.2005.00049.x}, 2005.

\bibitem[{\textit{{Knauth} and {Lowe}}(2003)}]{knauth03}
{Knauth}, L.~P., and D.~R. {Lowe}, {High Archean climatic temperature inferred
  from oxygen isotope geochemistry of cherts in the 3.5 Ga Swaziland
  Supergroup, South Africa}, \textit{Geological Society of America Bulletin},
  \textit{115}, 566--580,
  \doi{10.1130/0016-7606(2003)115<0566:HACTIF>2.0.CO;2}, 2003.

\bibitem[{\textit{{Krissansen-Totton} and
  {Catling}}(2017)}]{krissansen-totton17}
{Krissansen-Totton}, J., and D.~C. {Catling}, {Constraining climate sensitivity
  and continental versus seafloor weathering with an inverse geological carbon
  cycle model}, \textit{Nature Communications}, \textit{8},
  \doi{10.1038/ncomms15423}, 2017.

\bibitem[{\textit{{Le Hir} et~al.}(2008{\natexlab{a}})\textit{{Le Hir},
  {Godd{\'e}ris}, {Donnadieu}, and {Ramstein}}}]{lehir08a}
{Le Hir}, G., Y.~{Godd{\'e}ris}, Y.~{Donnadieu}, and G.~{Ramstein}, {A
  geochemical modelling study of the evolution of the chemical composition of
  seawater linked to a ''snowball'' glaciation}, \textit{Biogeosciences},
  \textit{5}, 253--267, 2008{\natexlab{a}}.

\bibitem[{\textit{{Le Hir} et~al.}(2008{\natexlab{b}})\textit{{Le Hir},
  {Ramstein}, {Donnadieu}, and {Godd{\'e}ris}}}]{lehir08b}
{Le Hir}, G., G.~{Ramstein}, Y.~{Donnadieu}, and Y.~{Godd{\'e}ris}, {Scenario
  for the evolution of atmospheric pCO$_{2}$ during a snowball Earth},
  \textit{Geology}, \textit{36}, 47--50, 2008{\natexlab{b}}.

\bibitem[{\textit{{Leconte} et~al.}(2013)\textit{{Leconte}, {Forget},
  {Charnay}, {Wordsworth}, and {Pottier}}}]{leconte13b}
{Leconte}, J., F.~{Forget}, B.~{Charnay}, R.~{Wordsworth}, and A.~{Pottier},
  {Increased insolation threshold for runaway greenhouse processes on
  Earth-like planets}, \textit{Nature}, \textit{504}, 268--271,
  \doi{10.1038/nature12827}, 2013.

\bibitem[{\textit{{Manabe} and {Bryan}}(1985)}]{manabe85}
{Manabe}, S., and K.~{Bryan}, Jr., {CO2-induced change in a coupled
  ocean-atmosphere model and its paleoclimatic implications}, \textit{Journal
  of Geophysical Research}, \textit{90}, 11, \doi{10.1029/JC090iC06p11689},
  1985.

\bibitem[{\textit{{Marin-Carbonne} et~al.}(2014)\textit{{Marin-Carbonne},
  {Chaussidon}, and {Robert}}}]{marin-carbonne14}
{Marin-Carbonne}, J., M.~{Chaussidon}, and F.~{Robert}, {The silicon and oxygen
  isotope compositions of Precambrian cherts: A record of oceanic
  paleo-temperatures?}, \textit{Precambrian Research}, \textit{247}, 223--234,
  \doi{10.1016/j.precamres.2014.03.016}, 2014.

\bibitem[{\textit{{Marty} et~al.}(2016)\textit{{Marty}, {Avice}, {Sano},
  {Altwegg}, {Balsiger}, {H{\"a}ssig}, {Morbidelli}, {Mousis}, and
  {Rubin}}}]{marty16}
{Marty}, B., G.~{Avice}, Y.~{Sano}, K.~{Altwegg}, H.~{Balsiger},
  M.~{H{\"a}ssig}, A.~{Morbidelli}, O.~{Mousis}, and M.~{Rubin}, {Origins of
  volatile elements (H, C, N, noble gases) on Earth and Mars in light of recent
  results from the ROSETTA cometary mission}, \textit{Earth and Planetary
  Science Letters}, \textit{441}, 91--102, \doi{10.1016/j.epsl.2016.02.031},
  2016.

\bibitem[{\textit{{Moore} and {Webb}}(2013)}]{moore13}
{Moore}, W.~B., and A.~G. {Webb}, {Heat-pipe Earth}, \textit{Nature},
  \textit{501}, 501--505, \doi{doi:10.1038/nature12473}, 2013.

\bibitem[{\textit{{Nisbet} and {Sleep}}(2001)}]{nisbet01}
{Nisbet}, E.~G., and N.~H. {Sleep}, {The habitat and nature of early life},
  \textit{Nature}, \textit{409}, 1083--1091, 2001.

\bibitem[{\textit{{Nutman} et~al.}(2015)\textit{{Nutman}, {Bennett}, and
  {Friend}}}]{nutman15}
{Nutman}, A.~P., V.~C. {Bennett}, and C.~R.~L. {Friend}, {The emergence of the
  Eoarchaean proto-arc: evolution of a c. 3700 Ma convergent plate boundary at
  Isua, southern West Greenland}, \textit{Geological Society of London Special
  Publications}, \textit{389}, 113--133, \doi{10.1144/SP389.5}, 2015.

\bibitem[{\textit{{Ohta} et~al.}(1996)\textit{{Ohta}, {Maruyama}, {Takahashi},
  {Watanabe}, and {Kato}}}]{ohta96}
{Ohta}, H., S.~{Maruyama}, E.~{Takahashi}, Y.~{Watanabe}, and Y.~{Kato}, {Field
  occurrence, geochemistry and petrogenesis of the Archean Mid-Oceanic Ridge
  Basalts (AMORBs) of the Cleaverville area, Pilbara Craton, Western
  Australia}, \textit{Lithos}, \textit{37}, 199--221,
  \doi{10.1016/0024-4937(95)00037-2}, 1996.

\bibitem[{\textit{Ozak et~al.}(2016)\textit{Ozak, Aharonson, and
  Halevy}}]{ozak16}
Ozak, N., O.~Aharonson, and I.~Halevy, Radiative transfer in co2-rich
  atmospheres: 1. collisional line mixing implies a colder early mars,
  \textit{Journal of Geophysical Research: Planets}, \textit{121}(6), 965--985,
  \doi{10.1002/2015JE004871}, 2015JE004871, 2016.

\bibitem[{\textit{{Pierazzo} et~al.}(1998)\textit{{Pierazzo}, {Kring}, and
  {Melosh}}}]{pierazzo98}
{Pierazzo}, E., D.~A. {Kring}, and H.~J. {Melosh}, {Hydrocode simulation of the
  Chicxulub impact event and the production of climatically active gases},
  \textit{Journal of Geophysical Research}, \textit{103}, 28,607--28,625, \doi{10.1029/98JE02496}, 1998.

\bibitem[{\textit{{Pope} et~al.}(2012)\textit{{Pope}, {Bird}, and
  {Rosing}}}]{pope12}
{Pope}, E.~C., D.~K. {Bird}, and M.~T. {Rosing}, {Isotope composition and
  volume of Earth√ïs early oceans}, \textit{Proceedings of the National
  Academy of Sciences}, \textit{109}(12), 4371--4376,
  \doi{10.1073/pnas.1115705109}, 2012.

\bibitem[{\textit{{Robert} and {Chaussidon}}(2006)}]{robert06}
{Robert}, F., and M.~{Chaussidon}, {A palaeotemperature curve for the
  Precambrian oceans based on silicon isotopes in cherts}, \textit{Nature},
  \textit{443}, 969--972, 2006.

\bibitem[{\textit{{Roberts} and {Spencer}}(2015)}]{roberts15}
{Roberts}, N., and C.~{Spencer}, {The zircon archive of continent formation
  through time}, \textit{Geological Society, London, Special Publications},
  \textit{389}, 197–225, 2015.

\bibitem[{\textit{Rosing et~al.}(2010)\textit{Rosing, Bird, Sleep, and
  Bjerrum}}]{rosing10}
Rosing, M.~T., D.~K. Bird, N.~H. Sleep, and C.~J. Bjerrum, {No climate paradox
  under the faint early Sun}, \textit{Nature}, \textit{464}(7289), 744--747,
  \doi{10.1038/nature08955}, 2010.

\bibitem[{\textit{{Rye} et~al.}(1995)\textit{{Rye}, {Kuo}, and
  {Holland}}}]{rye95}
{Rye}, R., P.~H. {Kuo}, and H.~D. {Holland}, {Atmospheric carbon dioxide
  concentrations before 2.2 billion years ago}, \textit{Nature}, \textit{378},
  603--605, \doi{10.1038/378603a0}, 1995.

\bibitem[{\textit{{Sacramento} and {Gruber}}(2006)}]{sarmiento06}
{Sacramento}, J., and N.~{Gruber}, \textit{{OCEAN BIOGEOCHEMICAL DYNAMICS}},
  2006.

\bibitem[{\textit{{Sheldon}}(2006)}]{sheldon06}
{Sheldon}, N., {Precambrian paleosols and atmospheric CO2 levels},
  \textit{Precambrian Research}, \textit{147}, 148--155,
  \doi{10.1016/j.precamres.2006.02.004}, 2006.

\bibitem[{\textit{{Shuvalov}}(2009)}]{shuvalov09}
{Shuvalov}, V., {Atmospheric erosion induced by oblique impacts},
  \textit{Meteoritics and Planetary Science}, \textit{44}, 1095--1105,
  \doi{10.1111/j.1945-5100.2009.tb01209.x}, 2009.

\bibitem[{\textit{{Sleep}}(2007)}]{sleep07}
{Sleep}, N.~H., {Plate Tectonics through Time}, \textit{Treatise on
  Geophysics,}, \textit{9}, 145--169, 2007.

\bibitem[{\textit{{Sleep} and {Zahnle}}(2001)}]{sleep01}
{Sleep}, N.~H., and K.~{Zahnle}, {Carbon dioxide cycling and implications for
  climate on ancient Earth}, \textit{Journal of Geophysical Research},
  \textit{106}, 1373--1400, \doi{10.1029/2000JE001247}, 2001.

\bibitem[{\textit{Stern}(2007)}]{stern07}
Stern, R.~J., {When and how did plate tectonics begin? Theoretical and
  empirical considerations}, \textit{Chinese Sci. Bull.}, \textit{52},
  578--591, 2007.

\bibitem[{\textit{{Tang} et~al.}(2016)\textit{{Tang}, {Chen}, and
  {Rudnick}}}]{tang16}
{Tang}, M., K.~{Chen}, and R.~L. {Rudnick}, {Archean upper crust transition
  from mafic to felsic marks the onset of plate tectonics}, \textit{Science},
  \textit{351}, 372--375, \doi{10.1126/science.aad5513}, 2016.

\bibitem[{\textit{{Tartese} et~al.}(2016)\textit{{Tartese}, M., {Gurenko},
  {Delarue}, and {Robert}}}]{tartese16}
{Tartese}, R., C.~M., A.~{Gurenko}, F.~{Delarue}, and F.~{Robert}, {Warm
  Archean oceans reconstructed from oxygen isotope composition of early-life
  remnants}, \textit{Geochemical Perspective letters}, \textit{3}, 55--65,
  \doi{10.1126/science.aad5513}, 2016.

\bibitem[{\textit{{van den Boorn} et~al.}(2007)\textit{{van den Boorn}, {van
  Bergen}, {Nijman}, and {Vroon}}}]{vandenboorn07}
{van den Boorn}, S., M.~{van Bergen}, W.~{Nijman}, and P.~{Vroon}, {Dual role
  of seawater and hydrothermal fluids in Early Archean chert formation:
  Evidence from silicon isotopes}, \textit{Geology}, \textit{35}, 939,
  \doi{10.1130$\%$2FG24096A.1}, 2007.

\bibitem[{\textit{{Viehmann} et~al.}(2014)\textit{{Viehmann}, {Hoffmann},
  {Munker}, and {Bau}}}]{viehmann14}
{Viehmann}, S., J.~E. {Hoffmann}, C.~{Munker}, and M.~{Bau}, {Decoupled Hf-Nd
  isotopes in Neoarchean seawater reveal weathering of emerged continents},
  \textit{Geology}, \textit{42}, 115--118, \doi{10.1130/G35014.1}, 2014.

\bibitem[{\textit{{Wolf} and {Toon}}(2013)}]{wolf13}
{Wolf}, E.~T., and O.~B. {Toon}, {Hospitable Archean Climates Simulated by a
  General Circulation Model}, \textit{Astrobiology}, \textit{13}, 656--673,
  \doi{10.1089/ast.2012.0936}, 2013.

\bibitem[{\textit{{Wordsworth} et~al.}(2017)\textit{{Wordsworth}, {Kalugina},
  {Lokshtanov}, {Vigasin}, {Ehlmann}, {Head}, {Sanders}, and
  {Wang}}}]{wordsworth17}
{Wordsworth}, R., Y.~{Kalugina}, S.~{Lokshtanov}, A.~{Vigasin}, B.~{Ehlmann},
  J.~{Head}, C.~{Sanders}, and H.~{Wang}, {Transient reducing greenhouse
  warming on early Mars}, \textit{Geophysical Research Letters}, \textit{44}, 665--671,
  \doi{10.1002/2016GL071766}, 2017.

\bibitem[{\textit{{Zahnle} and {Sleep}}(2002)}]{zahnle02}
{Zahnle}, K., and N.~H. {Sleep}, {Carbon dioxide cycling through the mantle and
  implications for the climate of ancient Earth}, \textit{Geological Society of
  London Special Publications}, \textit{199}, 231--257,
  \doi{10.1144/GSL.SP.2002.199.01.12}, 2002.

\end{thebibliography}







\end{document}